\documentclass[useAMS,usenatbib,fleqn]{mn2e}
\usepackage{graphicx}
\usepackage[abs]{overpic}
\usepackage[fleqn]{amsmath}
\usepackage[usenames,dvipsnames]{color}
\usepackage{times}

\usepackage{color}
\usepackage{xspace}

\title[Refining the Galaxy Merger Rate using Morphological Information]
{Galaxy And Mass Assembly (GAMA): Refining the Local Galaxy Merger Rate using Morphological Information}

\author[K. R. V. Casteels et al.]{Kevin R. V. Casteels,$^{1,2}$\thanks{E-mail: kcasteels@gmail.com (KRVC)}
Christopher J. Conselice,$^{1}$
Steven P. Bamford,$^{1}$
Eduard Salvador-Sol\'{e},$^{2}$\newauthor
Peder R. Norberg,$^{3}$
Nicola K. Agius,$^{4}$
Ivan Baldry,$^{5}$
Sarah Brough,$^{6}$
Michael J. I. Brown,$^{7}$\newauthor
Michael J. Drinkwater,$^{8}$
Simon P. Driver,$^{9,10}$
Alister W. Graham,$^{11}$
Joss Bland-Hawthorn,$^{12}$\newauthor
Andrew M. Hopkins,$^{6}$
Lee S. Kelvin,$^{9,10,13}$
Angel R. L\'{o}pez-S\'{a}nchez,$^{6,14}$
Jon Loveday,$^{15}$\newauthor
Aaron S. G. Robotham,$^{9,10}$
Jos\'{e} A. V\'{a}zquez-Mata,$^{15}$
\smallskip\\
$^{1}$School of Physics and Astronomy, The University of Nottingham, University Park, Nottingham, NG7 2RD, UK \\
$^{2}$Institut de Ci\`{e}ncies del Cosmos, Universitat de Barcelona, Mart\'{i} Franqu\`{e}s 1, E-08028 Barcelona, Spain \\
$^{3}$Institute for Computational Cosmology, Department of Physics, Durham University, South Road, Durham DH1 3LE, UK \\
$^{4}$Jeremiah Horrocks Institute, University of Central Lancashire, Preston PR1 2HE, UK \\
$^{5}$Astrophysics Research Institute, Liverpool John Moores University, IC2, Liverpool Science Park, 146 Brownlow Hill, Liverpool, L3 5RF, UK\\
$^{6}$Australian Astronomical Observatory, PO Box 915, North Ryde, NSW 1670, Australia \\
$^{7}$School of Physics, Monash University, Clayton, Victoria 3800, Australia \\
$^{8}$Department of Physics, University of Queensland, Brisbane, Queensland 4072, Australia \\
$^{9}$International Centre for Radio Astronomy Research (ICRAR), University of Western Australia, Crawley, WA 6009, Australia \\
$^{10}$Scottish Universities Physics Alliance (SUPA), School of Physics and Astronomy, University of St Andrews, North Haugh, St Andrews KY16 9SS, UK \\
$^{11}$Centre for Astrophysics and Supercomputing, Swinburne University of Technology, Hawthorn, Victoria 3122, Australia \\
$^{12}$Sydney Institute for Astronomy, University of Sydney A28, NSW 2006, Australia \\
$^{13}$Institut fur Astro- und Teilchenphysik, Universitat Innsbruck, Technikerstrasse 25, 6020 Innsbruck, Austria \\
$^{14}$Department of Physics and Astronomy, Macquarie University, NSW 2109, Australia \\
$^{15}$Astronomy Centre, University of Sussex, Falmer, Brighton BN1 9QH, UK}
\begin{document}

\maketitle \begin{abstract}
We use the Galaxy And Mass Assembly (GAMA) survey to measure the local Universe mass dependent merger fraction and merger rate using galaxy pairs and the CAS structural method, which identifies highly asymmetric merger candidate galaxies.  
Our goals are to determine which types of mergers produce highly asymmetrical galaxies, and to provide a new measurement of the local galaxy major merger rate.  
We examine galaxy pairs at stellar mass limits down to $M_{*} = 10^{8}M_{\odot}$ with mass ratios of $<$100:1 and line of sight velocity differences of $\Delta V<500$ km s$^{-1}$.  We find a significant increase in mean asymmetries for projected separations less than the sum of the individual galaxy's Petrosian 90 radii.  For systems in major merger pairs with mass ratios of $<$4:1 both galaxies in the pair show a strong increase in asymmetry, while  in minor merger systems (with mass ratios of $>$4:1) the lower mass companion becomes highly asymmetric, while the larger galaxy is much less affected.  
The fraction of highly asymmetric paired galaxies which have a major merger companion is highest for the most massive galaxies and drops progressively with decreasing mass.  
We calculate that the mass dependent major merger fraction is fairly constant at $\sim1.3-2\%$ between $10^{9.5}<M_{*}<10^{11.5} M_{\odot}$, and increases to $\sim4\%$ at lower masses. When the observability time scales are taken into consideration, the major merger rate is found to approximately triple over the mass range we consider.  The total co-moving volume major merger rate over the range $10^{8.0}<M_{*}<10^{11.5} M_{\odot}$ is $(1.2 \pm 0.5) \times 10^{-3}$ $h^{3}_{70}$ Mpc$^{-3}$ Gyr$^{-1}$.  
\end{abstract}


\pagerange{\pageref{firstpage}--\pageref{lastpage}} \pubyear{2013}

\clearpage
\label{firstpage}

\section{Introduction}

The most accepted theories of galaxy and structure formation state that galaxies are formed hierarchically, where smaller galaxies merge to form larger galaxies (e.g. \citealt{whit1978,kauf1993,cole2000}).  Essentially all of the galaxies we observe in the local universe are the products of mergers or have been influenced by interactions with other galaxies. 
The mass ratio of the merging progenitor galaxies is known to have a major effect on the merger remnant.  To distinguish between the different types of mergers, those taking place between progenitor galaxies with similar masses are referred to as major mergers (mass ratio of $<$4:1), while mergers between galaxies with large mass ratios are called minor mergers (mass ratio of $>$4:1).

The strongly fluctuating gravitational forces and shock waves experienced during violent relaxation in major mergers are known to funnel gas into their cores and induce intense star-bursts and feed central black holes as seen both in simulations and in observations \citep{barn1991,barn1996,miho1994,miho1996,dima2005,hopk2005a,hopk2005b, cox2008,lotz2008, lotz2010a, lotz2010b, dima2012, elli2013, torr2014}.   This results in a rapid exhaustion of the gas supply and can lead to the formation of elliptical galaxies, as well as efficiently creates the spheroidal components of galaxies (e.g. \citealt{hopk2010b}).  Understanding the mass dependence of the major merger rate can help to tell us how the red sequence and the spheroidal components of galaxies are being built up over different mass ranges.

According to the predictions of merger rates of dark matter halos (e.g. \citealt{fakh2008,fakh2010}), minor mergers are expected to be much more common than major mergers.  Although there is not a direct correlation between halo mergers and the galaxies they contain, the same is expected to be true for galaxies.  
Indeed  papers such as \citet{lotz2011, lope2011,bluc2012} and \citet{kavi2014} find the galaxy minor merger rate to be several times higher than the major merger rate.  The higher frequency of minor mergers plays an important role in the gradual build up of massive disk galaxies, without the total destruction of the primary galaxy, as is the case in major mergers.  \citet{hopk2010b} showed that minor mergers can also build up pre-existing bulges in massive galaxies, as well as form new bulges in lower mass galaxies.  It is possible that minor mergers may sometimes induce bar structures \citep{skib2011}, while there is evidence that major mergers have the opposite effect, and destroy bars \citep{mend2011,lee2012,cast2012}.

Accurately measuring the mass dependence of galaxy merger rates is important for fully understanding the phenomenon described above, as well as providing a way to test hierarchical galaxy formation models and simulations (e.g. \citealt{bens2002}).  The merger history is known to increase as a function of redshift, often very steeply (e.g. \citealt{lefe2000,patt2002,cons2003b,cons2008,lin2004,lotz2008,joge2009,lope2011,tasc2014}), such that the merger process is a dominant one in the formation of at least the most massive galaxies (e.g. \citealt{cons2006,lope2011,tasc2014}).  By understanding and having a reliable value for the nearby mass dependent merger rate we can calibrate the increase in the merger rate for galaxies seen at higher redshifts to obtain a full picture of the role of merging in galaxies.

Previous attempts to measure the mass dependent merger fraction using studies of close pairs have found it to be constant or increase slightly with mass (e.g. \citealt{xu2004,patt2008,domi2009,xu2012}).  On the other hand, the study of \citet{brid2010} which selected interacting and merging galaxies based on their morphologies found evidence for a mildly decreasing interaction fraction with mass.   
However, these studies may be measuring different aspects of the merger process, as the \citet{brid2010} sample also includes remnants of mergers identified from tidal features, and not exclusively interacting pairs.  While many studies have attempted to measure the nearby galaxy merger fraction and overall merger rate (e.g. \citealt{depr2007}), the actual \emph{mass dependent} merger rate is not yet known with much accuracy.  Along with this study, two companion papers by \citet{depr2014} and \citet{robo2014} also use GAMA data to study the luminosity and mass dependent galaxy merger rate using close pairs.

In this study we use the CAS (Concentration, Asymmetry and Smoothness) method of \citet{cons2003b} to identify highly asymmetric galaxies as merger candidates, and determine which of these are produced in major mergers (through the examination of close pairs) to obtain a measurement of the major merger fraction and rate for nearby galaxies.  
Galaxies are known to become highly asymmetric in major mergers, and sometimes also in minor mergers if the progenitor cool gas fractions are high enough \citep{lotz2010b}.
Galaxies in very close pairs often have high asymmetries due to the strong tidal forces at small separations \citep{hern2005, patt2005,depr2007}, and measuring the mass ratios of highly asymmetric pairs provides us with a way to estimate the contribution of both major and minor mergers to the population of morphologically disturbed and highly asymmetric galaxies, and thus obtain a clean sample of major mergers in the nearby universe.

In this paper we present the observed merger fractions for galaxies in the nearby universe as a function of stellar mass, as well as projected separation for those which are in galaxy pairs.  We determine the proportion of highly asymmetric galaxies produced in major mergers compared to those produced in minor mergers.  We then utilize this to obtain a measurement of the merger fraction and rate for nearby galaxies as a function of stellar mass, as well as the total merger rate in the nearby universe.

This paper is structured as follows: in Section \ref{gamadata} we describe the data set and sample selection,  in Section \ref{gamaresults} we present our method and results, and in Section \ref{gamadiscussion} we discuss their implications and summarize our conclusions in Section \ref{gamaconclusions}.
A $\Lambda CDM$ cosmology is assumed throughout, with $\Omega_{\Lambda}$ = 0.7, $\Omega_{m}$ = 0.3, and 
H$_0 = 70$ km s$^{-1}$ Mpc$^{-1}$.

\section{Data and Sample Selection}\label{gamadata}

\subsection{GAMA}

Galaxy and Mass Assembly (GAMA; \citealt{driv2009,driv2011}) is a galaxy survey covering $\sim 300$ degrees$^2$ of sky down to $r \sim 19.8$ mag.  Galaxy spectra have been obtained by the 3.9-m Anglo-Australian Telescope using the AAOmega multi-object spectrograph. For the brighter galaxies, GAMA also uses existing spectra from the Sloan Digital Sky Survey (SDSS; \citealt{york2000}), the 2dF Galaxy Redshift Survey (2dFGRS; \citealt{coll2001}) and the Millennium Galaxy Catalogue (MGC; \citealt{lisk2003}). 
The GAMA I NGP sample \citep{bald2010,driv2011} used in this study consists of 114 094 SDSS selected galaxies and is $\sim99\%$ spectroscopically complete \citep{hopk2013}.  This sample consists of three equatorial regions of $\sim 48$ degrees$^2$ each, two of which are complete for $r<19.4$ and one which is complete for $r<19.8$.  This high spectroscopic completeness makes GAMA the ideal survey to study galaxies in close pairs, as there is essentially no dependence of incompleteness on angular separation, as there is in the MGC and SDSS. The photometrically derived galaxy stellar mass estimates of \citet{tayl2011} are used in this study.  The photometry used to determine the masses are based on an improved reanalysis of SDSS imaging \citep{hill2011}.  Not all galaxies have mass estimates, due to poor photometric measurements, and so the overall completeness of the sample used here drops slightly to $\sim98\%$.

The sample used here to calculate the mass dependent merger fraction and merger rate consists of 51,700 galaxies with stellar masses  $M_{*} > 10^{6.5} M_{\odot}$ and redshifts $0.001<z<0.2$ (median $z = 0.129$).  
The maximum redshift limit of $z_{max}=0.2$ is chosen to ensure that all of the galaxies are sufficiently resolved to provide robust CAS measurements. 
Over this redshift range the sample is approximately $\sim 95\%$ stellar mass complete for $M_{*} > 10^{10} M_{\odot}$ (\citealt{tayl2011}).  
The final sample contains 1470 highly asymmetric galaxies (defined in section \ref{CASdefs}), 142 of which are found to have a very close companion.

\subsection{CAS Measurements}\label{CASdefs}

The CAS method \citep{cons2000,cons2003b} was used to obtain quantitative measurements of the morphological properties of the galaxies. The CAS parametrization consists of three measurements: the galaxy concentration index $C$, an asymmetry index $A$ and a clumpiness index $S$. The parameters $A$ and $S$ are particularly useful for identifying merging galaxies in the later stages of a merger.   How these are measured is defined below.

The parameter $A$ is a measure of the rotational symmetry of a galaxy, and is obtained by rotating its image 180 degrees and subtracting the light within 1.5$\times$ the Petrosian $\eta = 0.2$ radius from the original monochromatic fits image.  The value of the asymmetry is such that more asymmetric systems have a higher value of $A$, with $A > 0.35$ typically for merging systems. The centre for rotation is decided
by an iterative process which finds the location of the minimum asymmetry.  The parameter $A$ is the value measured when taking the ratio of the subtracted flux to the original galaxy flux, and is given by

\begin{equation}\label{aeq}
A=\rm{min} \left( \sum |I_{0} - I_{180}| \over \sum|I_{0}| \right) - \rm{min} \left( \sum|B_{0} - B_{180}| \over \sum |I_{0}| \right) ,
\end{equation}
where $I_{0}$ is the intensity of the original image, $I_{180}$ is the intensity of the rotated image, and $B_{0}$ and $B_{180}$ are noise corrections obtained by iteratively repeating the same rotation and subtraction on empty background regions.
The value of $A$ can range from 0 to 2, where $A=0$ represents a completely symmetrical galaxy, and $A=2$ represents a completely asymmetrical galaxy.

The parameter $S$ is a measure of highly localized, bright structures, and is meant to detect bright star forming regions.
$S$ is defined as the ratio of the amount of light in high spatial frequency structures within 1.5$\times$ the Petrosian 90 radius, to the total amount of light within that radius \citep{cons2003b}.  The Petrosian 90 radius is defined as the radius which contains 90\% of the Petrosian flux $F_p$ \citep{petr1976}. To obtain a measure of the high frequency structure, a boxcar-smoothed image is produced from the original image.  $S$ is defined as follows,

\begin{equation}\label{seq}
S= 10 \left[ {\sum^{N,N}_{x,y=1,1} (I_{x,y} - I^{s}_{x,y}) \over \sum^{N,N}_{x,y=1,1} I_{x,y}} - {\sum^{N,N}_{x,y=1,1} B^{S}_{x,y} \over \sum^{N,N}_{x,y=1,1} I_{x,y}} \right] ,
\end{equation}
where $I_{x,y}$ is the intensity of light in a given pixel, $I^{s}_{x,y}$ is the intensity of that pixel in the image smoothed by $0.3r_{Pet}$, and $B_{x,y}$ is an intensity value of a pixel from a smoothed background region.

Merger candidate galaxies are then identified by selecting galaxies with $A>A_{limit}$ and by requiring $A>S$ to exclude galaxies with bright star-forming regions which do not have global asymmetries characteristic of a merger.  Throughout this paper $A_{limit}=0.35$ (e.g. \citealt{cons2003b}).  For a detailed description of the CAS method used in this paper the reader is referred to \citet{cons2003b}.

\subsubsection{Segmentation Maps}

Soon after beginning work with the GAMA data it became clear that there was a problem with the publicly available SDSS segmentation maps, which are meant to cleanly separate galaxies from each other and define their boundary. It was found that for galaxies with close angular separations, the public segmentation maps were consistently joining clearly separate galaxies together. For studies of the general population this may not be a big problem, but when dealing with merging galaxies it is essential that very close and clearly distinct galaxy pairs are cleanly separated.  Defining the correct boundaries of a galaxy is integral to obtaining a physically meaningful CAS measurement.  For example, if two galaxies share the same segmentation map (i.e. they have not been separated) then the resulting asymmetry measurement ($A$) will be artificially high regardless of the true asymmetry of the galaxies.  

Therefore the SExtractor code of \citet{bert1996} was used to create new segmentation maps for all of the GAMA galaxies.
Segmentation maps and subsequent CAS measurements were obtained for the SDSS DR7 \emph{r}-band images.  A random sample of 100 galaxies with small angular separation were selected to test our method, and the SExtractor parameters were adjusted to optimize the separation of these galaxies. Specifically the following parameters were changed from their default settings: DETECT\_MINAREA = 10, DETECT\_THRESH = 1.8, and ANALYSIS\_THRESH = 1.8 .

\begin{figure*}

\begin{tabular}{c c c c}
  \begin{overpic}[width=40mm,angle=270]{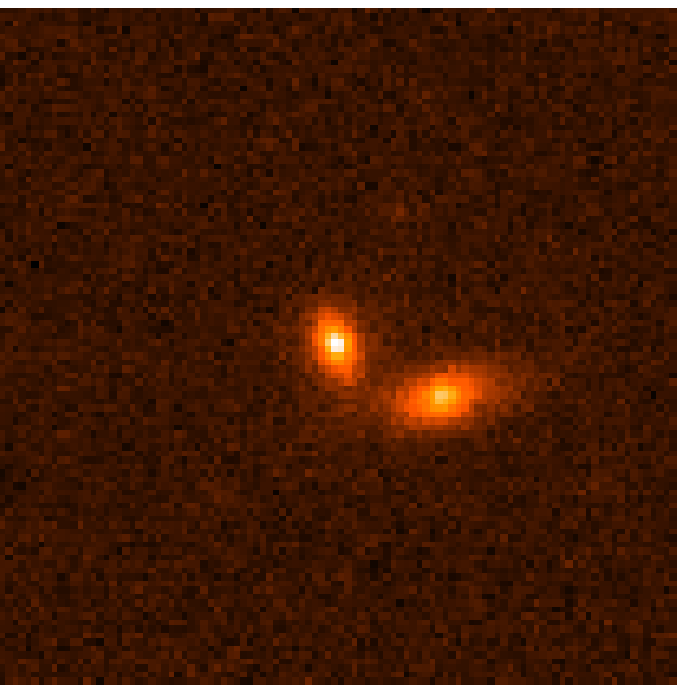}
	\put(24,100){\Large {\color{SkyBlue} \textbf{SDSS \emph{r}-band}}}
  \end{overpic}&
  \begin{overpic}[width=40mm,angle=270]{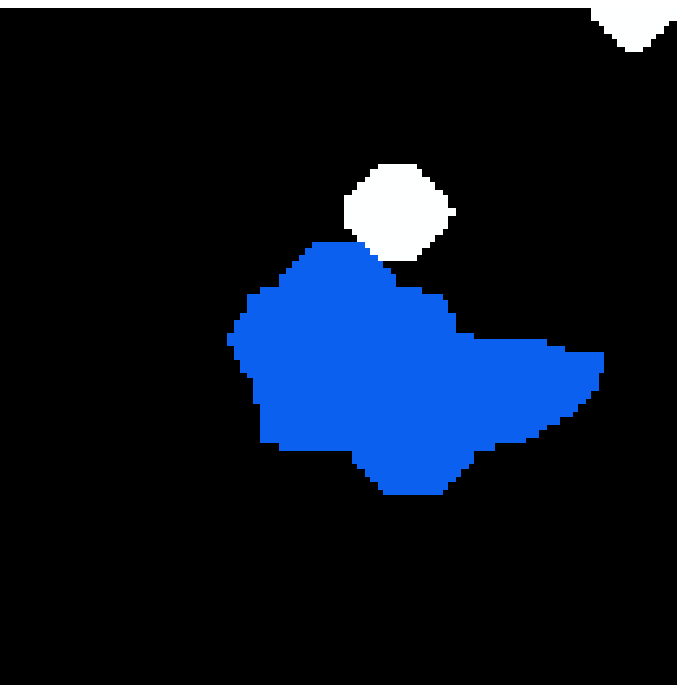}
	\put(30,100){\Large {\color{Green} \textbf{SDSS Map}}}
  \end{overpic}&
  \begin{overpic}[width=40mm,angle=270]{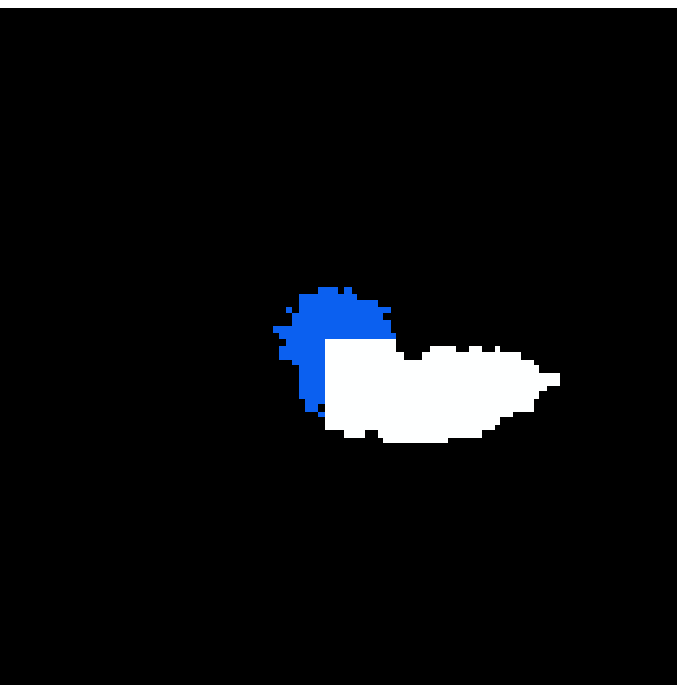}
	\put(18,100){\Large {\color{Orange} \textbf{SExtractor Map}}}
  \end{overpic}&
  \begin{overpic}[width=40mm,angle=270]{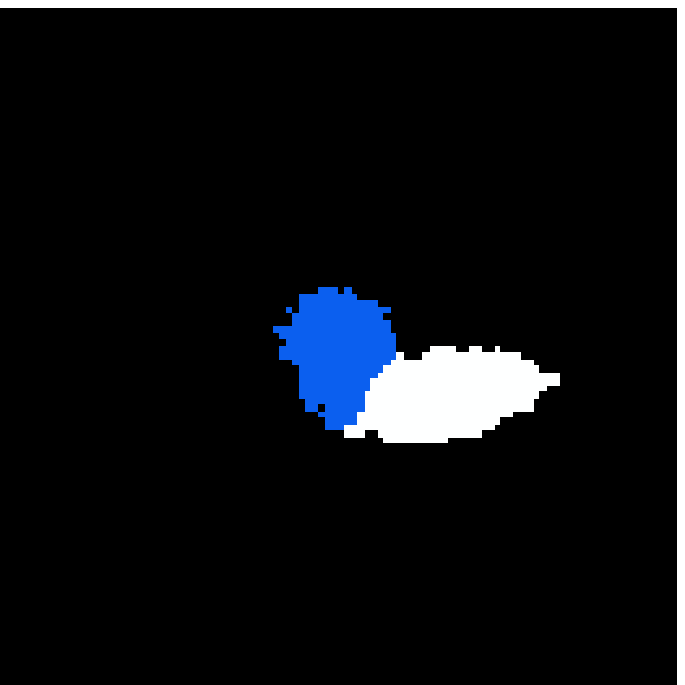}
	\put(24,100){\Large {\color{Red} \textbf{Manual Map}}}
  \end{overpic}\\
\end{tabular}
\caption{Example of a SDSS r band image compared to the different segmentation maps.  The public SDSS map covers two clearly separate galaxies with a single segmentation profile.  The raw SExtractor map does a better job of separating the galaxies, but the profile of the lower galaxy partly covers the centre of the upper galaxy, leading to an non-physical CAS measurement.  The manually corrected segmentation map realistically separates the galaxies and results in a correspondingly robust CAS measurement.}\label{segmaps}
\end{figure*}

To ensure the robustness of our segmentation maps and the resultant CAS measurements, the maps of all galaxies identified as being highly asymmetric ($A>0.35 , A>S$) were visually examined.  Out of 1455 highly asymmetric galaxies, 116 were found to have bad segmentation maps, mostly due to poor deblending of galaxies with small angular separations.  The GNU Image Manipulation Program (GIMP) was used to manually edit the problem segmentation maps pixel by pixel to ensure they accurately mapped out the target galaxies.  CAS measurements were then obtained for galaxies with the corrected segmentation maps.  Of the 116 potentially highly asymmetric galaxies with bad segmentations maps, 26 continued to have $A>0.35$ and $A>S$ with the corrected maps. Example images are shown in Figure \ref{segmaps}, comparing an original SDSS map, an initial SExtractor map, and a manually corrected map.

The photometry used for the stellar mass estimates \citep{tayl2011,hill2011} makes use of SExtractor’s auto photometry. These segmentation maps have not been manually examined or corrected although the resultant photometry should not be nearly as affected by bad maps as CAS measurements are.  In most cases, even when close pairs are not properly deblended the galaxy centre will be correctly identified.  The photometry is based on a flexible, elliptical aperture and as long as the centre and approximate light distribution are known, the resultant photometry should be reasonable.

\section{Method and Results}\label{gamaresults}

\subsection{Definitions}\label{definitions}

The main objective of this study is to estimate the mass dependent galaxy merger rate.  To do so we first need to measure the number of highly asymmetric merger candidate galaxies ($N_{A}$) as a function of mass as follows

\begin{equation}\label{fsumeq}
N_{A} = \sum_{i = 1}^{N_{T}} { n_i(A>0.35; A>S)} ,
\end{equation}
where $N_{T}$ is the total number of galaxies in each stellar mass bin, with $n_i(A>0.35; A>S)$ = 1 if a given galaxy meets the criteria $A>0.35$ and $A>S$, and 0 otherwise.  Galaxies with $A>0.35$ and $A>S$ are referred to as highly asymmetric, while those that do not meet this criteria are referred to as non-asymmetric.  The fraction of all highly asymmetric galaxies as function of mass ($f_{asym}$) is found as follows

\begin{equation}\label{Asumeq}
 f_{asym} = \frac{N_{A}}{N_{T}} .
\end{equation}

\noindent The actual number of major merger galaxies ($N_{m}$) is found as follows

\begin{equation}\label{ftmeq}
  N_{m} = N_{A} \frac{f_{tm}f_{A_{4:1}}}{f_{am}},
\end{equation}
where $f_{tm}$ is the fraction of highly asymmetric galaxies which are truly mergers, $f_{A_{4:1}}$ is the fraction of highly asymmetric galaxies which are caused by major mergers with mass ratios $<$4:1, and $f_{am}$ is the fraction of mergers which become highly asymmetric during the merger process.  $f_{tm}$ is determined by visually examining all of the merger candidate galaxies and is discussed in Section \ref{Vchecksection}, while $f_{A_{4:1}}$ is measured in Section \ref{majorminorsep}.  In this work we assume all galaxies become highly asymmetric at some point in the merger process, so $f_{am} = 1$.  Although it is possible that in some high mass ratio mergers the low mass companion will become highly asymmetric while the high mass companion will not, we statistically correct the merger fraction to select only major mergers, so the assumption that $f_{am} = 1$ is still reasonable.

Following the definition in \cite{cons2006}, the fraction of galaxies in major mergers is given by

\begin{equation}\label{fgmeq}
  f_{major} = \frac{N_{m} \kappa }{N_{T}  + (\kappa - 1)N_{m} },
\end{equation}
where $\kappa$ is the average number of galaxies which merged to produce $N_{m}$.  In \citet{cons2006} it was argued that $\kappa$ must be $\ge 2$, but that is not necessarily true.  As we will show in Section \ref{majorminorsep}, a small fraction of highly asymmetric galaxies are found in pairs, and in situations where both paired galaxies are highly asymmetric, $\kappa < 2$.  Thus, we define $\kappa$ as follows:

\begin{equation}\label{kappaeq}
  \kappa = 2 - N_{P_{AA}}/N_{A},
\end{equation}
where $N_{P_{AA}}$ is the number of paired galaxies where both members are highly asymmetric.
The merger rate per galaxy ($R_{merger}$) can then be calculated as

\begin{equation}\label{majormergerrateeq}
R_{major} = \frac{f_{major}} {T_{merger,A}} ,
\end{equation}
where $T_{merger,A}$ is the time scale over which merging galaxies are observed to have $A>0.35$ and $A>S$.
As we will discuss further in Section \ref{castimescales}, $T_{merger,A}$ is a strong function of the mass ratios, masses and gas fractions of the merging galaxies.  The co-moving volume merger rate is defined as

\begin{equation}\label{comovemergerrateeq}
\Gamma_{major} = \frac{f_{major} \times \phi} {T_{merger,A}} ,
\end{equation}
where $\phi$ is the co-moving number density of galaxies in a given mass bin.  Note that this is inverse of the notation used in \citet{cons2006}, but the same as \citet{hopk2010b}.

\subsection{Mass Dependent Asymmetry Fraction}

\begin{figure}
\begin{center}
  \begin{overpic}[width=2.2in,angle=270]{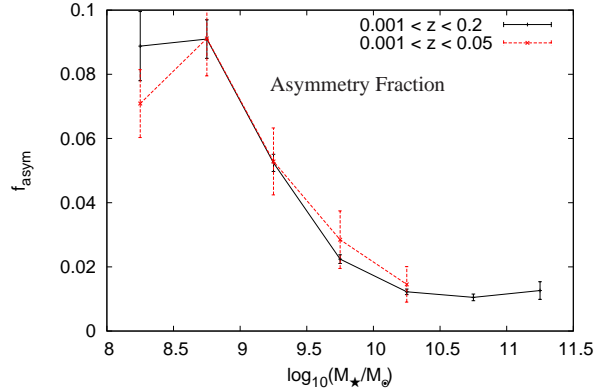}
	\put(100,120){Asymmetry Fraction}
  \end{overpic}
\end{center}
\caption{The mass dependent asymmetry fraction ($f_{asym}$) is plotted for two different redshift ranges.  The solid black line represents $f_{asym}$ for $0.001<z<0.2$ and the red dashed line represents $f_{asym}$ for $0.001<z<0.05$.  No highly asymmetric galaxies were detected for $M_{*}>10^{10.5} M_{\odot}$ in the $0.001<z<0.05$ range sample.}\label{A035}
\end{figure}

In Figure \ref{A035} the fraction of galaxies which are asymmetric, $f_{asym}$, is shown as a function of mass for systems with $10^{8.0}<M_{*}<10^{11.5} M_{\odot}$.  Between $10^{10.0}<M_{*}<10^{11.5} M_{\odot}$ we see that $f_{asym}$ is constant at $\sim1.2\%$ but for $M_{*}<10^{10.0} M_{\odot}$ the fraction increases significantly, up to $\sim9\%$ for $10^{8.0}<M_{*}<10^{9.0} M_{\odot}$.  The errors given for the mean asymmetry measurements are standard error in the mean, while jackknife errors are given for asymmetry fraction measurements.

The complete sample used in this study ($0.001<z<0.2$) begins to become significantly mass incomplete for the reddest galaxies with $M_{*} < 10^{10.0} M_{\odot}$, which is also the mass where the strong increase in the merger fraction begins (black solid line in Figure \ref{A035}).  
At the same time, red, low mass galaxies are much less common than blue, low mass galaxies, so it is not immediately clear how much of an effect this mass incompleteness has on the measurement of the asymmetry fraction.

To test this we select a sample of galaxies which is $\sim95\%$ mass complete for $M_{*} > 10^{8.0} M_{\odot}$ by restricting the redshift range to $0.001 < z < 0.05$ (\citealt{tayl2011}) resulting in a sample of 5066 galaxies.  The red dashed line in Figure \ref{A035} represents the fraction of galaxies with $A>0.35$ and $A>S$ for this smaller sample as a function of mass.  Despite the smaller sample size, it is clear that the fraction of highly asymmetric galaxies increases strongly for $M_{*} < 10^{10.0} M_{\odot}$, indicating that the mass incompleteness of the $0.001<z<0.2$ sample is not significantly contributing to the increased fraction of highly asymmetric low mass galaxies.

\subsection{Visual checking for non-merger systems}\label{Vchecksection}

\begin{figure*}

\begin{tabular}{c c c c}
	{\Large {\color{Blue} \textbf{Merger}}}&&&\\
	\includegraphics[width=40mm,angle=270]{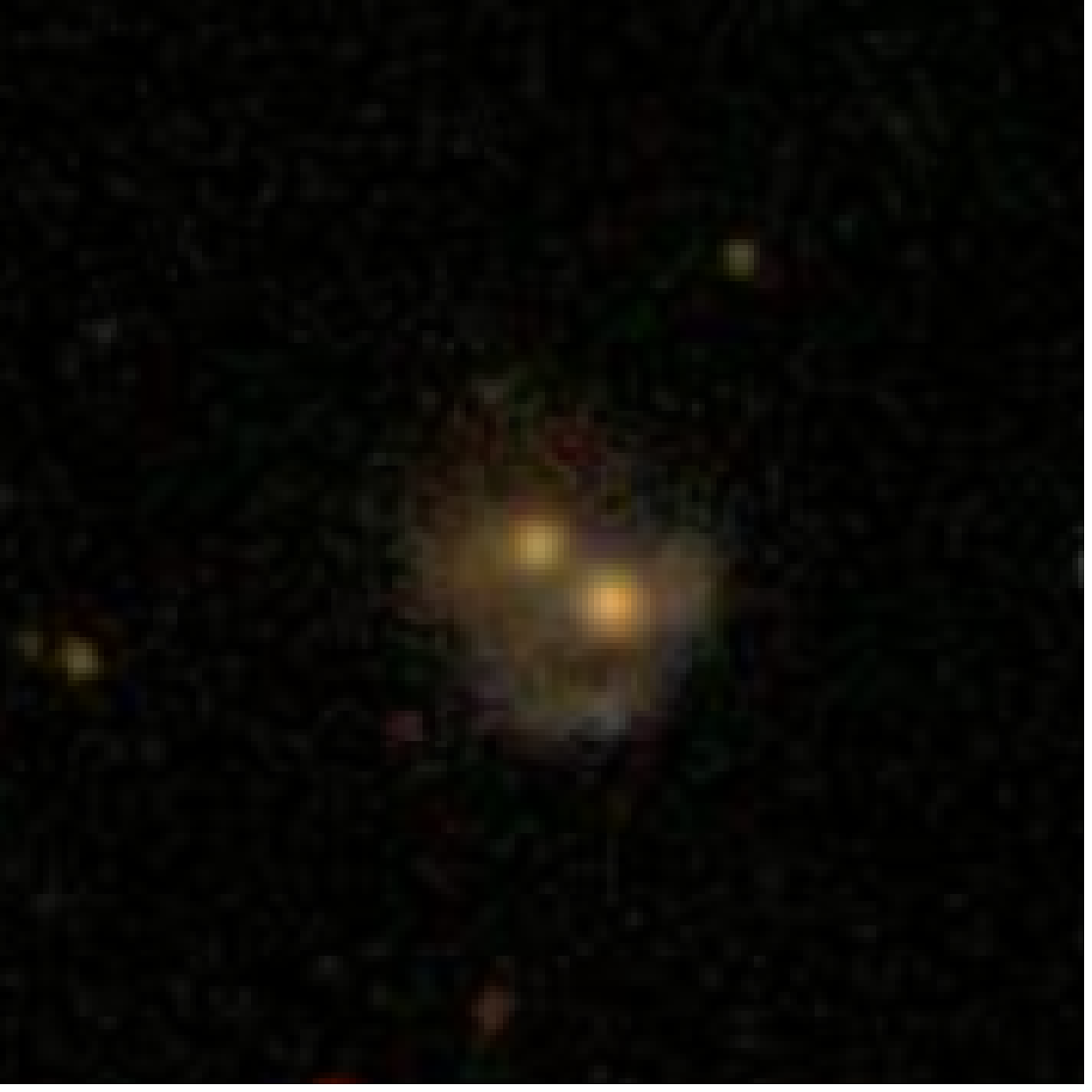}&
	\includegraphics[width=40mm,angle=270]{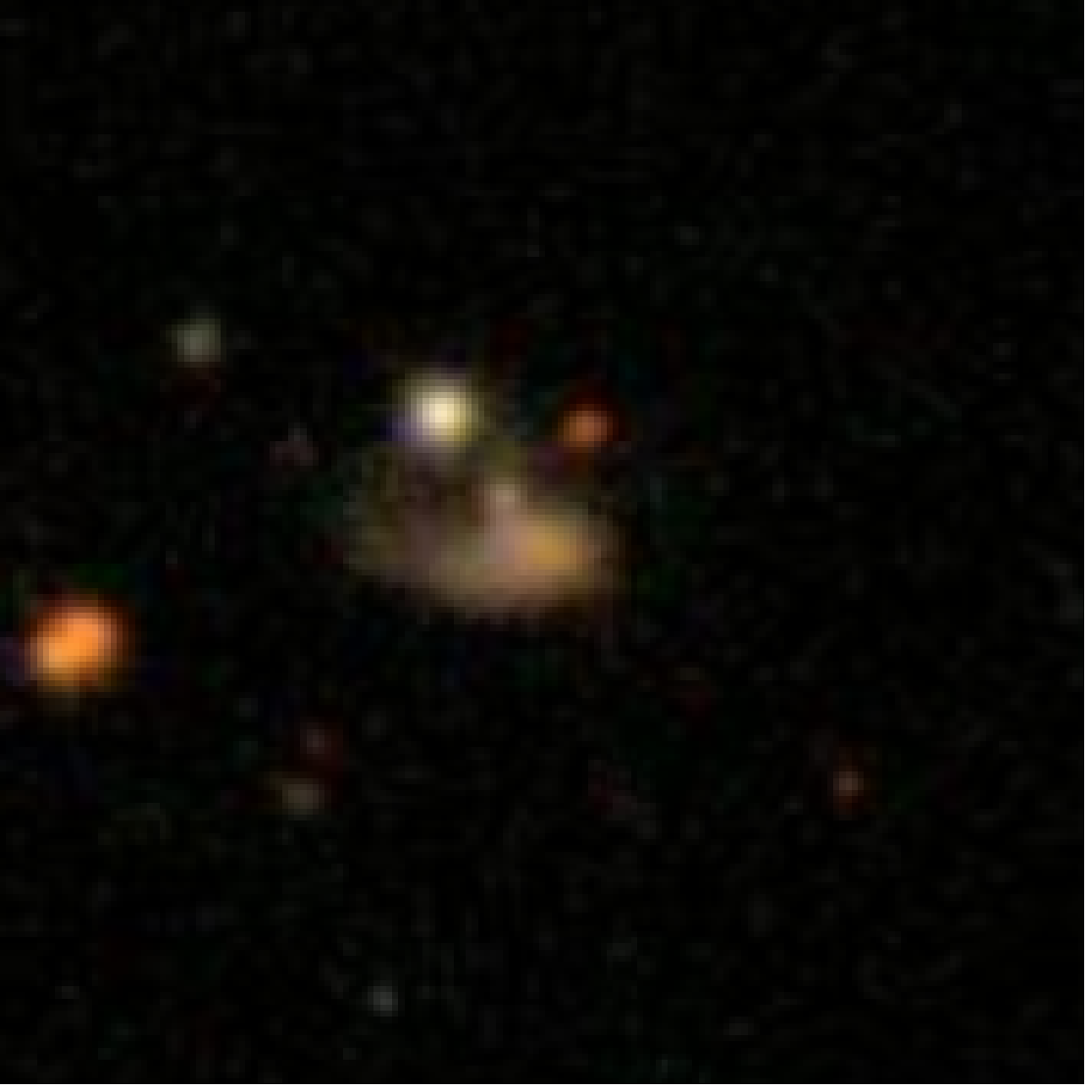}&
	\includegraphics[width=40mm,angle=270]{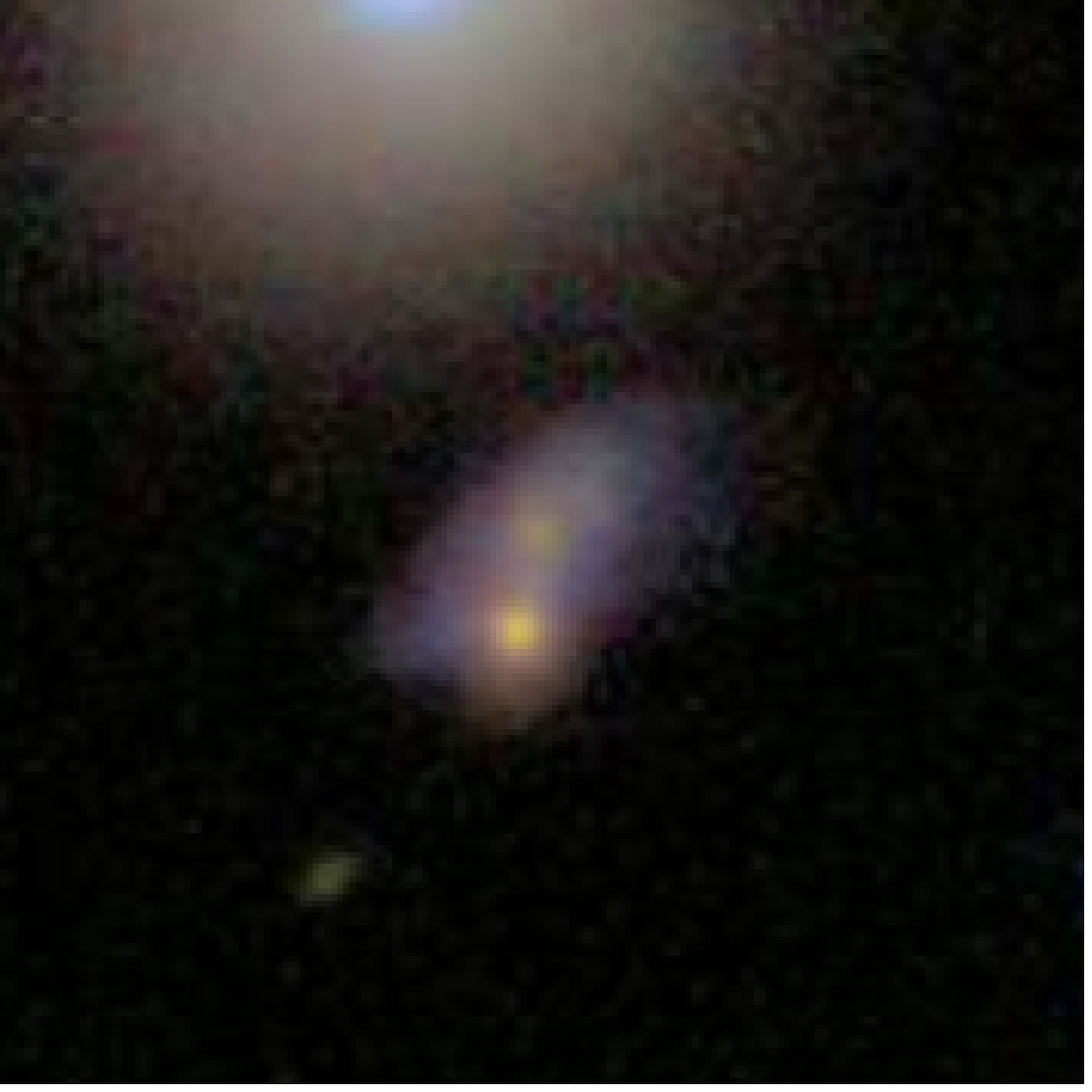}&
	\includegraphics[width=40mm,angle=270]{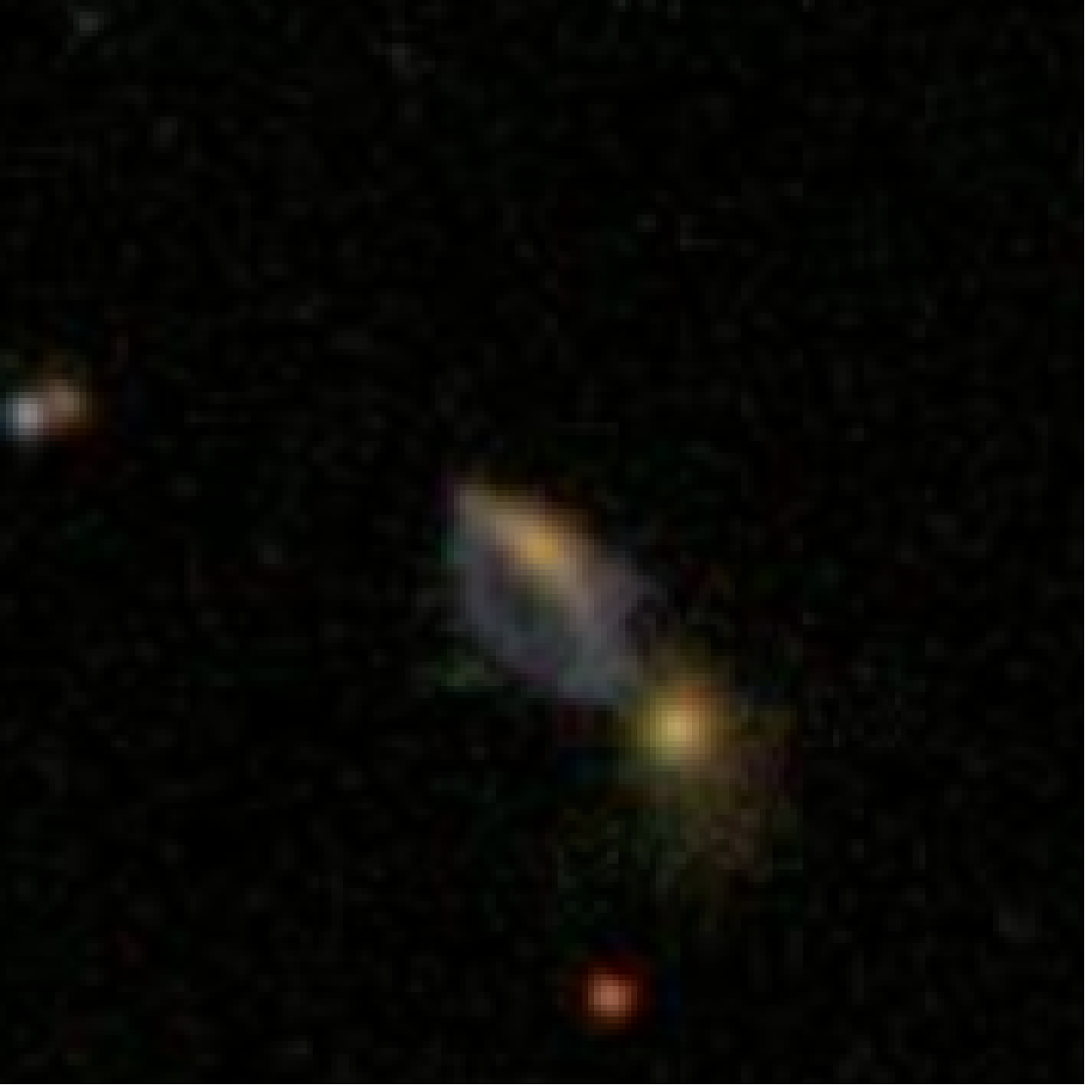}\\
	&&&\\
	{\Large {\color{Green} \textbf{Maybe}}}&&&\\
  	\includegraphics[width=40mm,angle=270]{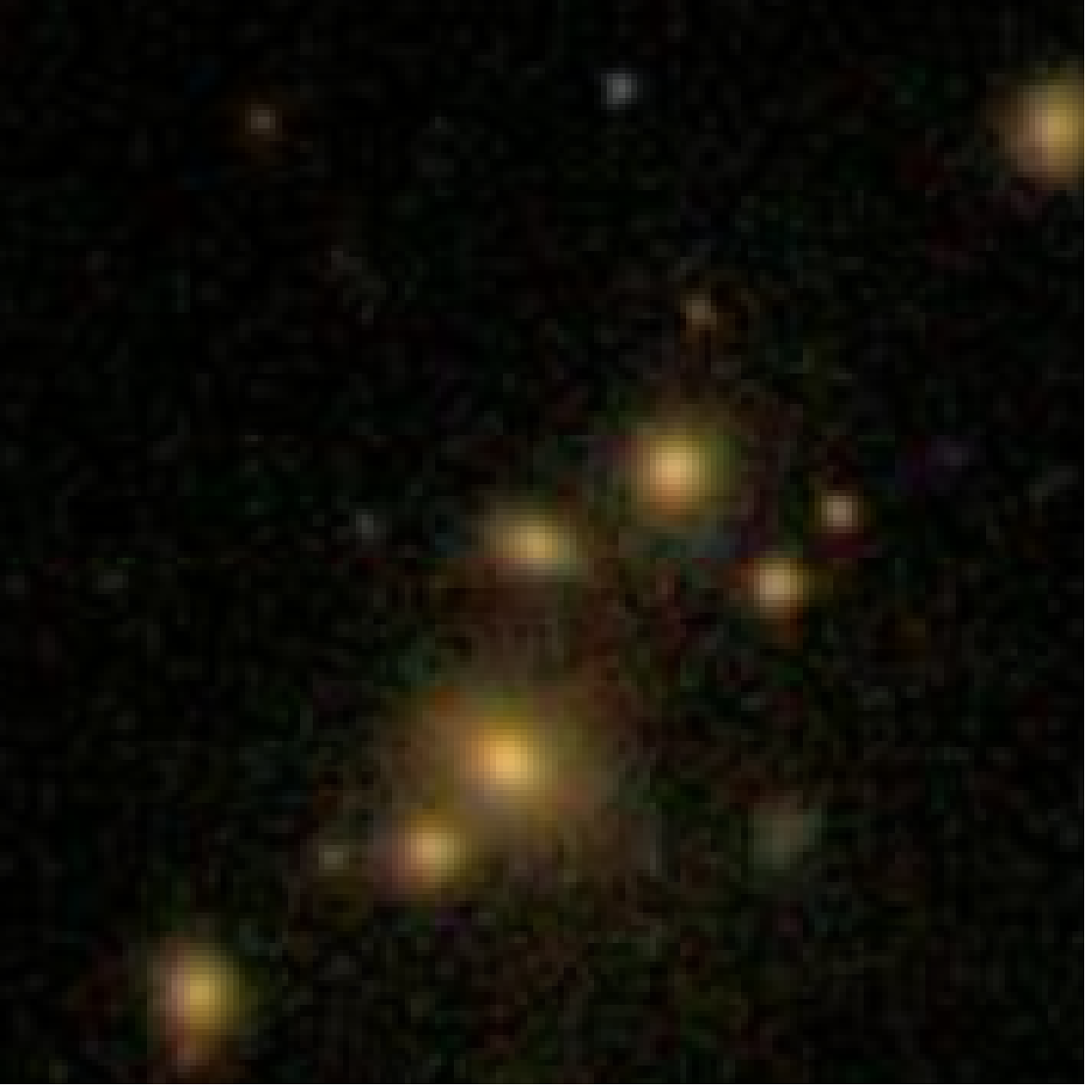}&
	\includegraphics[width=40mm,angle=270]{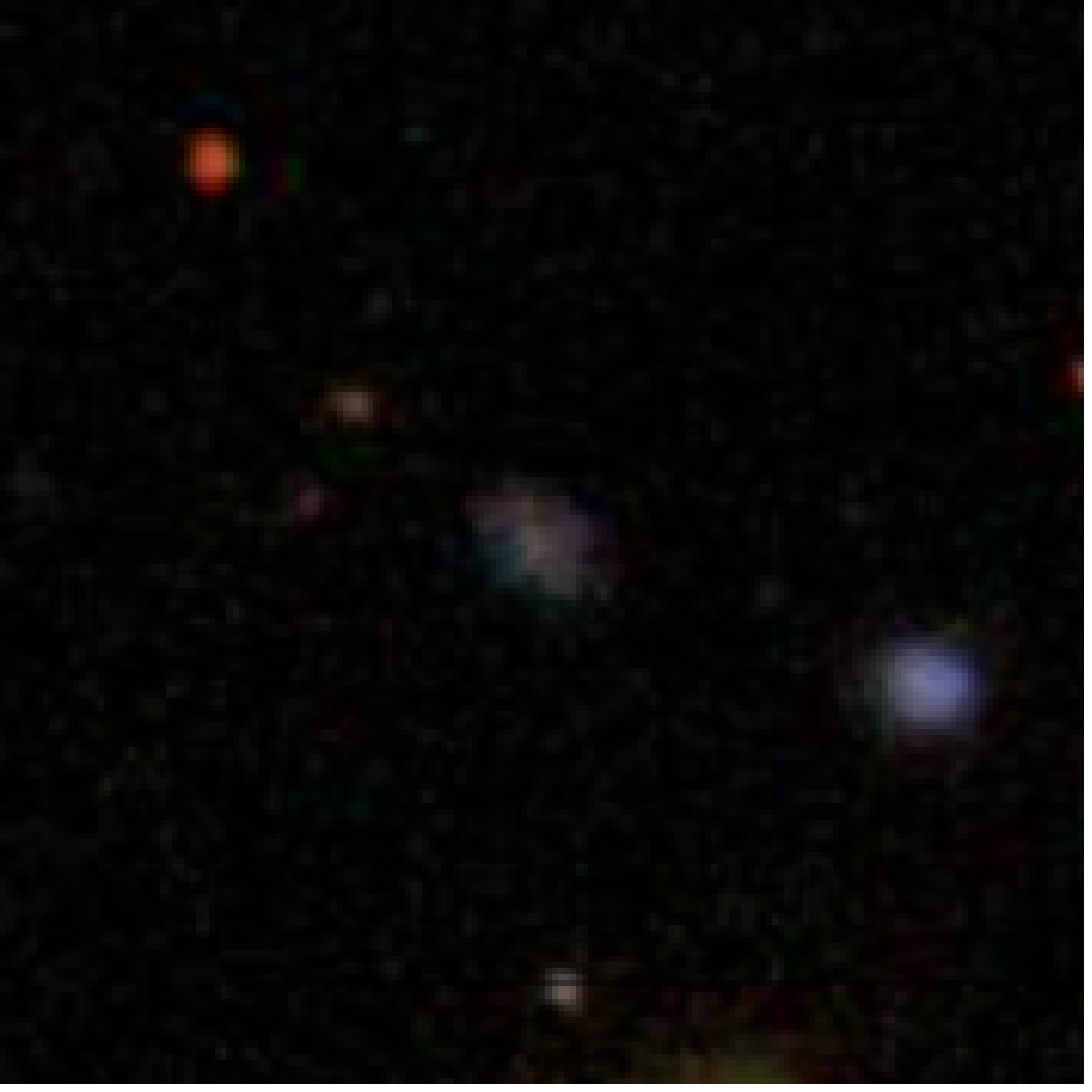}&
	\includegraphics[width=40mm,angle=270]{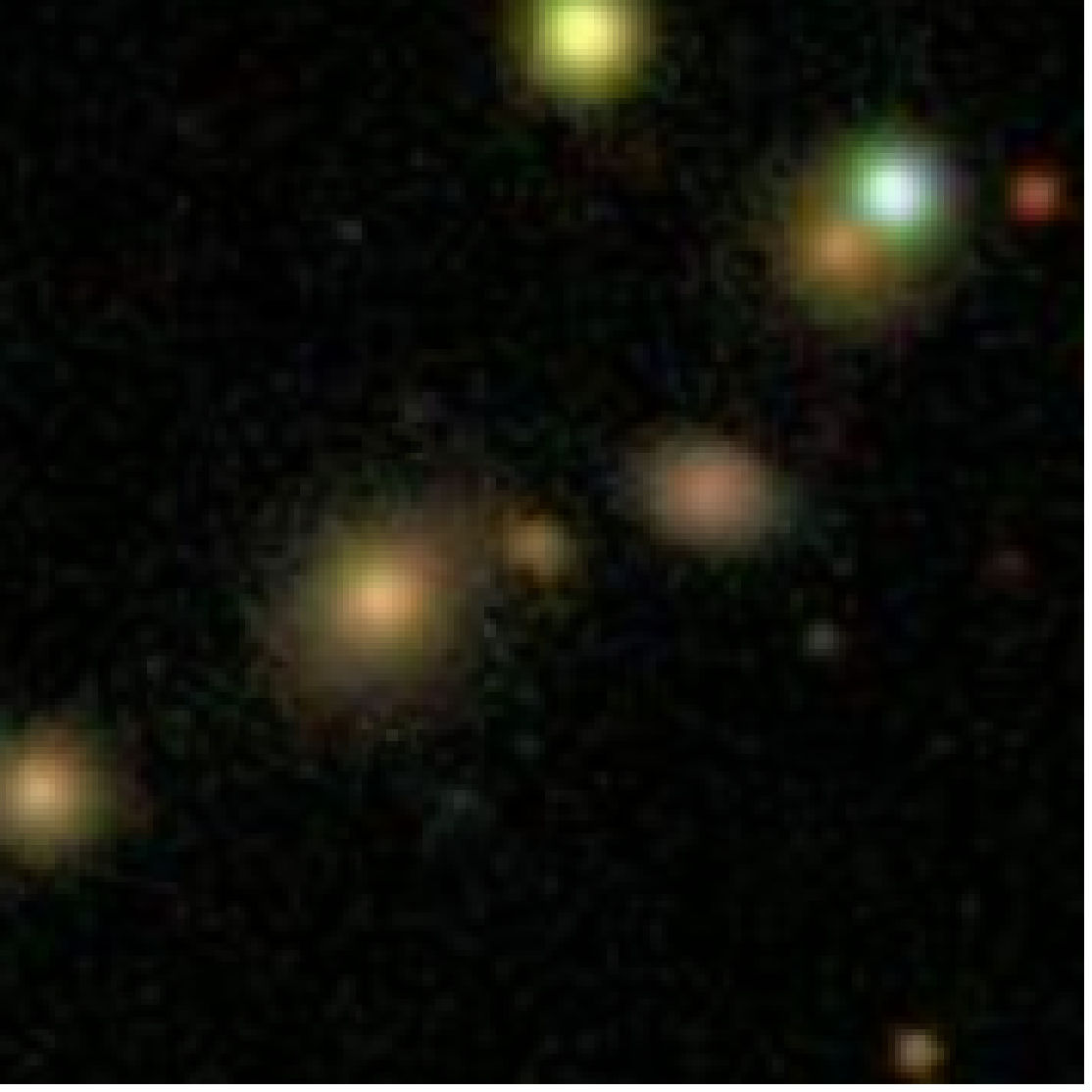}&
	\includegraphics[width=40mm,angle=270]{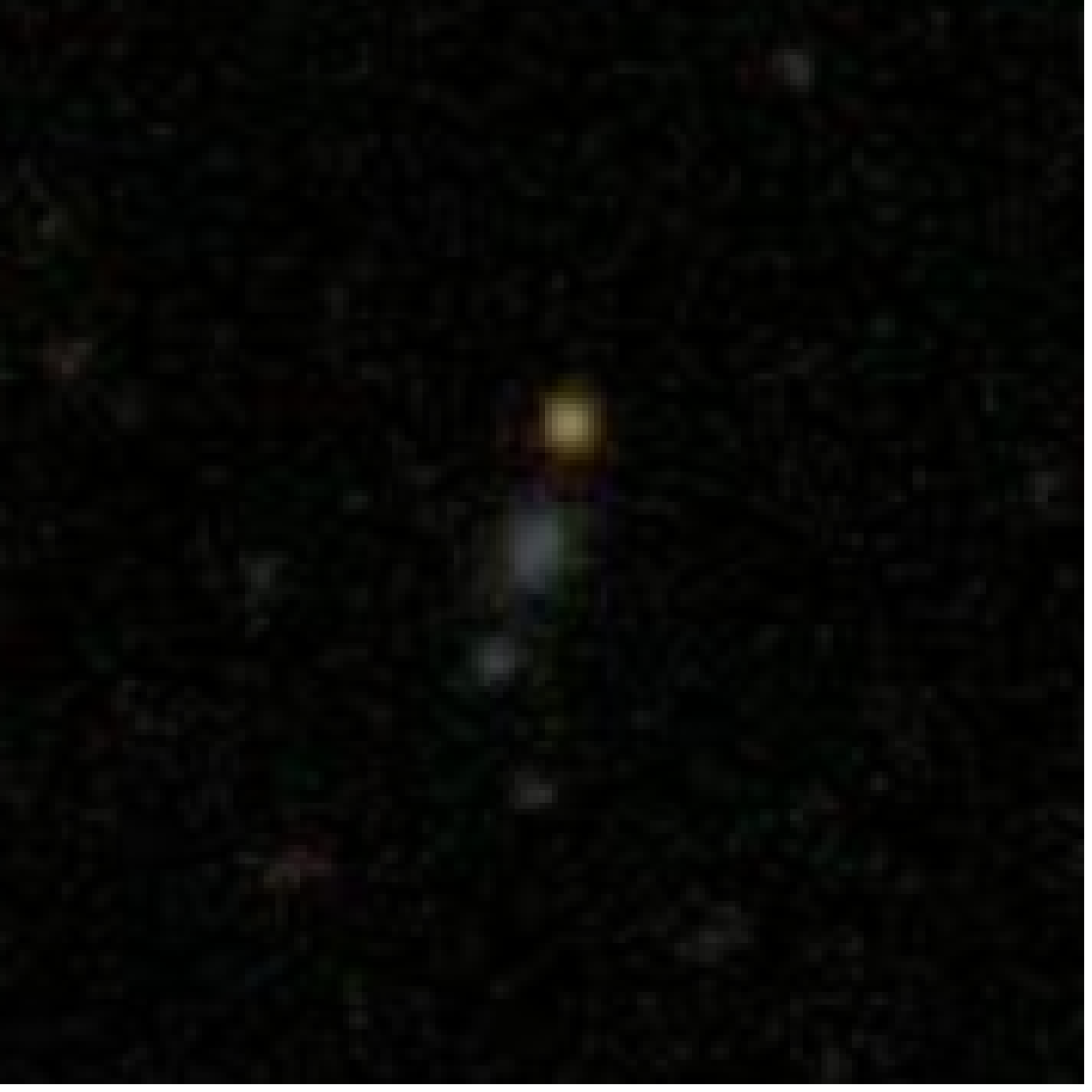}\\
	&&&\\
	{\Large {\color{Red} \textbf{Non-Merger}}}&&&\\
  	\includegraphics[width=40mm,angle=270]{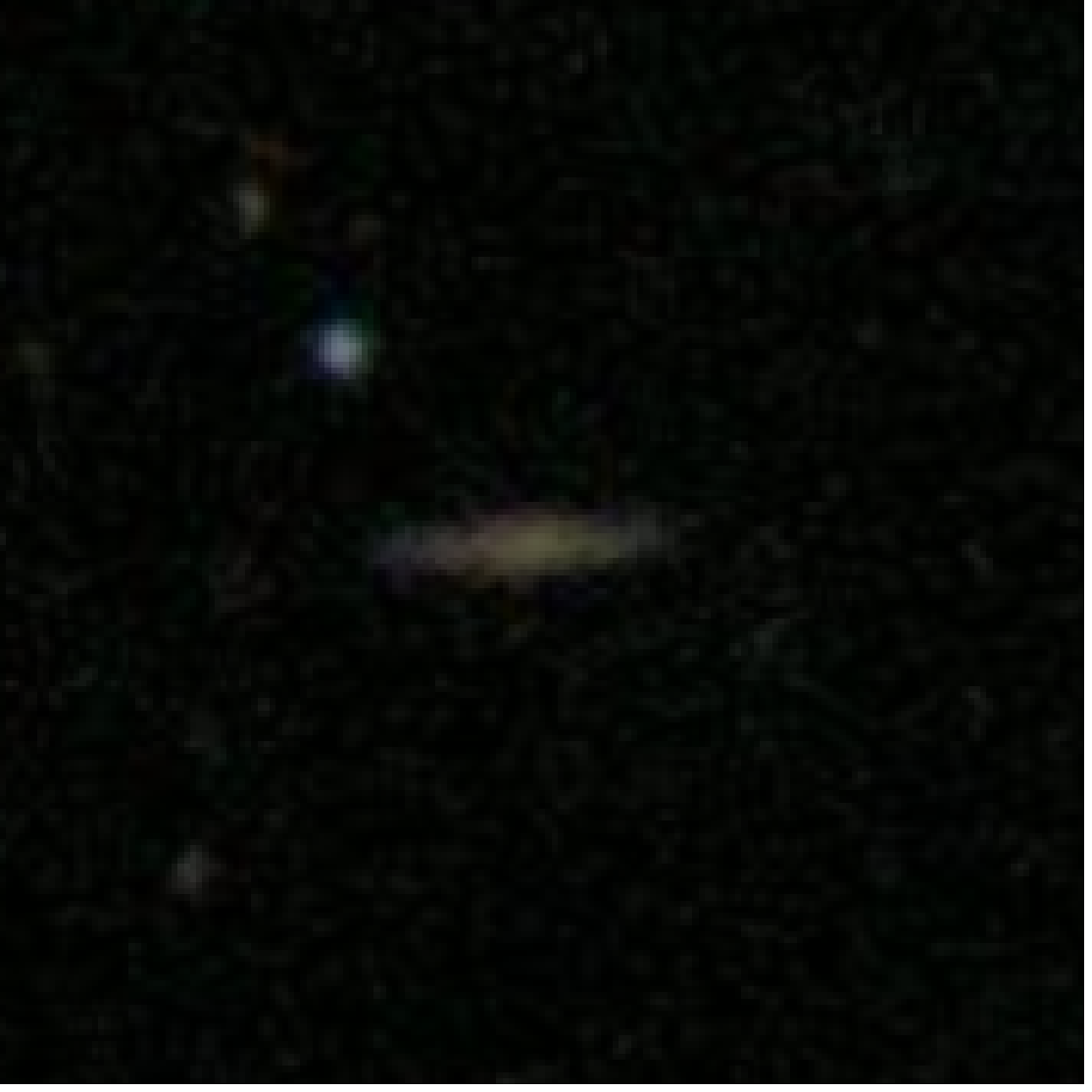}&
	\includegraphics[width=40mm,angle=270]{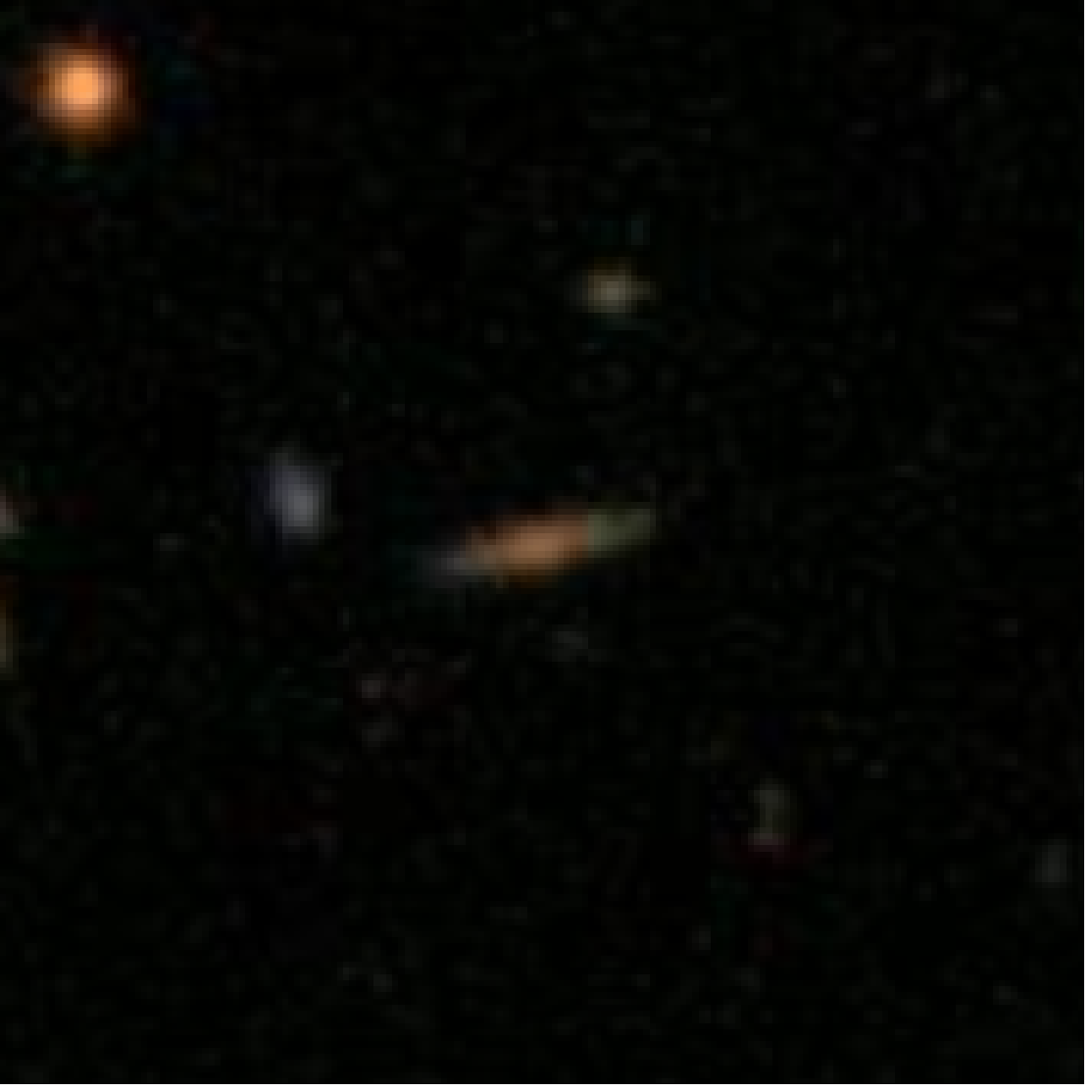}&
	\includegraphics[width=40mm,angle=270]{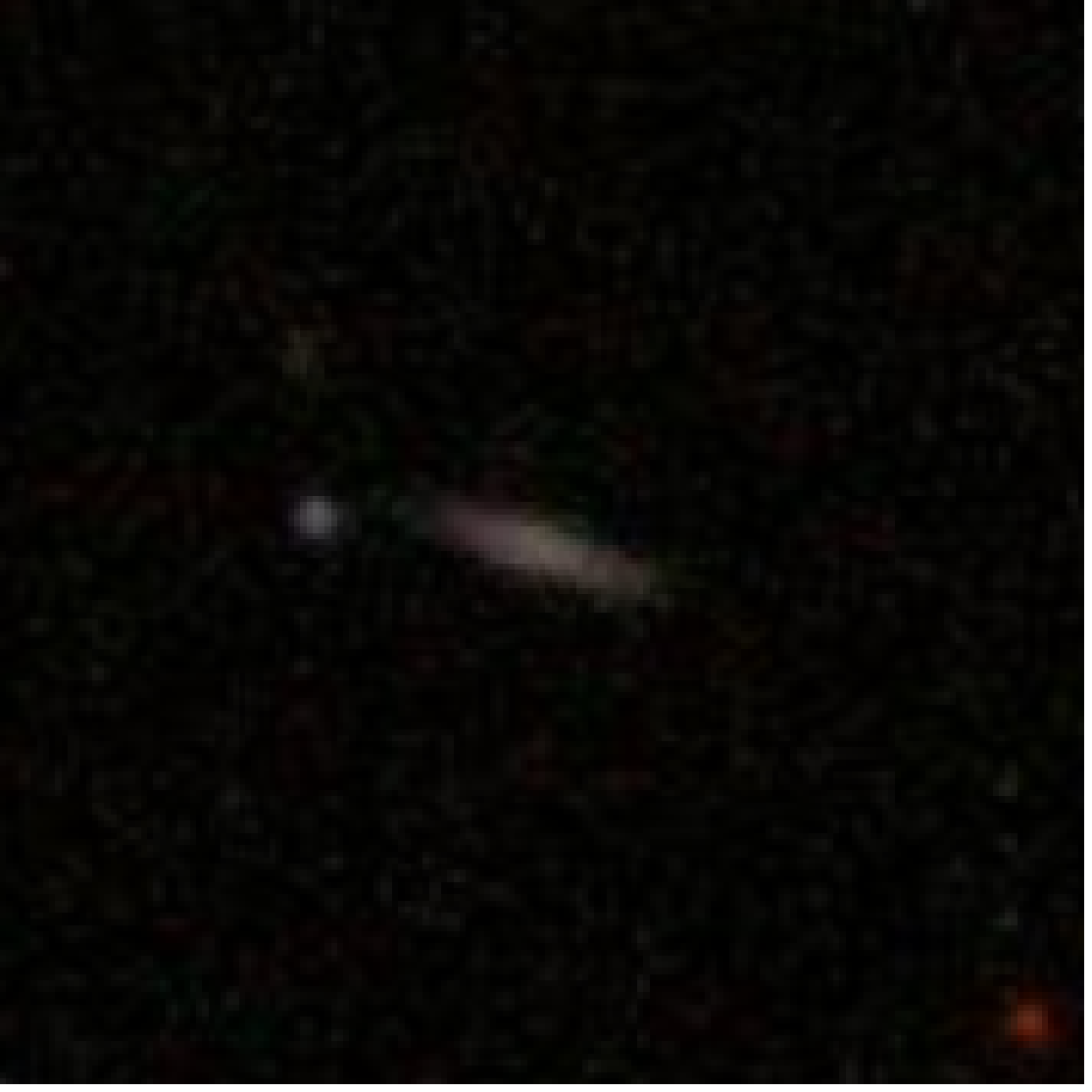}&
	\includegraphics[width=40mm,angle=270]{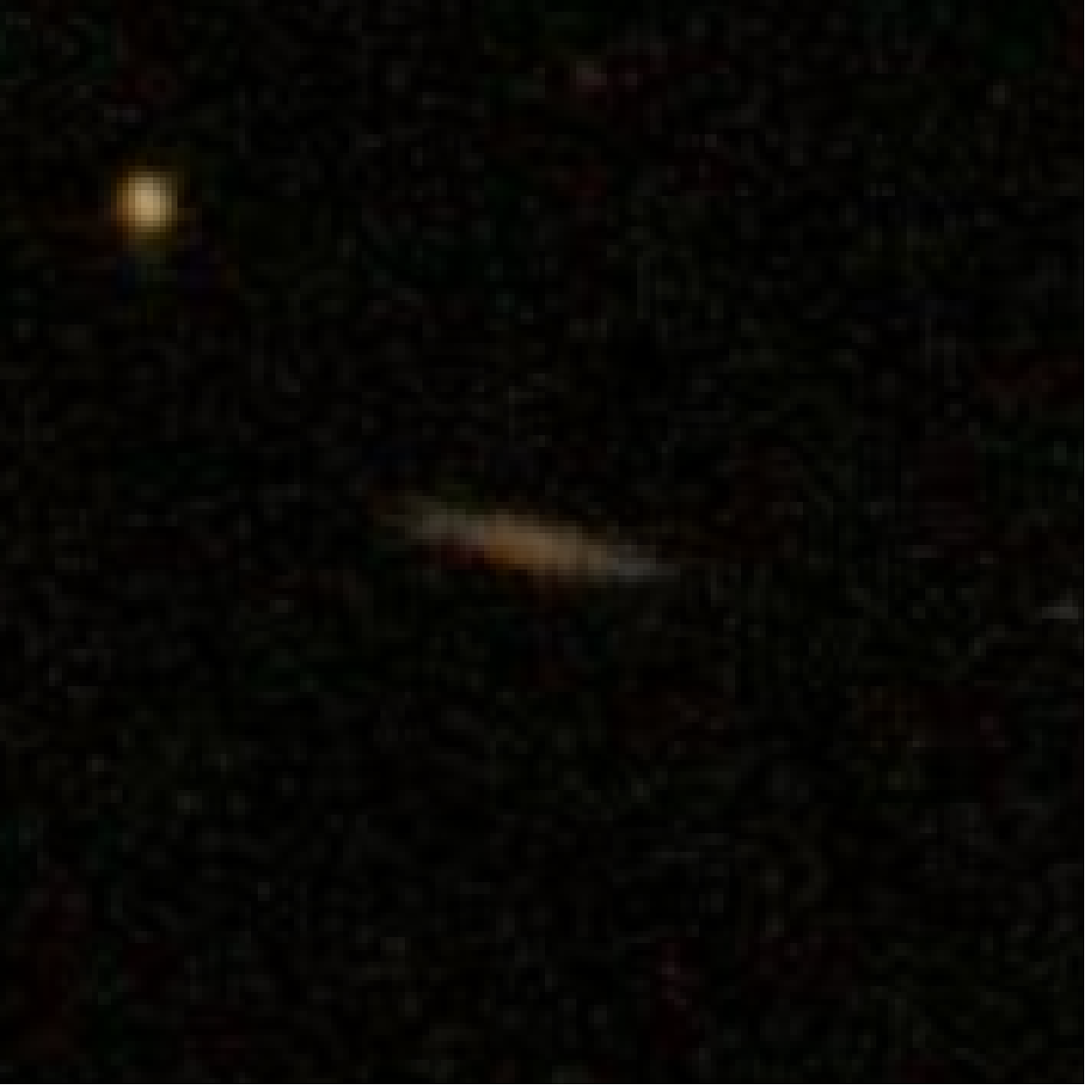}\\
\end{tabular}
\caption{Examples of galaxies classified as \emph{Merger}, \emph{Maybe} and \emph{Non-merger}.  The galaxy of concern is the one in the centre of each image.}\label{maybemerger}
\end{figure*}

\begin{table}
\caption{Values for the visual checked highly asymmetric galaxies.  Each galaxies was classified as either \emph{Merger}, \emph{Maybe} or \emph{Non-Merger}.  The fraction of highly asymmetric galaxies visually confirmed to be mergers is given by $f_{tm}$ which is defined in Equation \ref{vunseq}.}\label{vcheckedtable}
\centering
\begin{tabular}{c c c c c c}
\hline\hline
$log(M_{*}/M_{\odot})$ & $N_{A}$ & $N_{\emph{Merger}}$ & $N_{\emph{Maybe}}$ & $N_{\emph{Non-Merg}}$ &  $f_{tm}$ \\
\hline
8.25   & 72   &	 25   &	 41   & 6  & 0.63$\pm$0.06 \\
8.75   & 248  &	 90	  &  142  &	16 & 0.65$\pm$0.03 \\
9.25   & 399  &	 167  &	 209  &	23 & 0.68$\pm$0.02 \\
9.75   & 279  &	 135  &	 119  &	25 & 0.70$\pm$0.03 \\
10.25  & 201  &	 117  &	 66	  &	18 & 0.75$\pm$0.03 \\
10.75  & 102  &	 72	  &  24	  &	6  & 0.82$\pm$0.04 \\
11.25  & 19   &	 14	  &  5    &	0  & 0.87$\pm$0.08 \\
\hline\hline
\end{tabular}
\end{table}

All galaxies with $A>0.35$ and $A>S$ were visually examined to determine the contamination of non-merger systems to $f_{asym}(M_*)$, which we define as $f_{tm}$ (Section \ref{definitions}).
Otherwise undisturbed galaxies can be scattered to higher asymmetries by image noise and artefacts, edge on galaxies with dust lanes creating apparent breaks and asymmetries, or a foreground star overlapping a galaxy's light profile, often leading to unrealistic segmentation maps and a high asymmetry measurement.

\begin{figure}
\begin{center}
  \begin{overpic}[width=2.2in,angle=270]{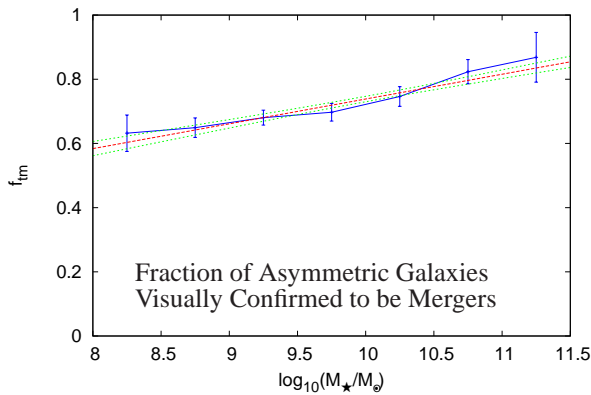}
	\put(50,50){\large Fraction of Asymmetric Galaxies} 
	\put(50,40){\large Visually Confirmed to be Mergers}
  \end{overpic}
\end{center}
\caption{The fraction of galaxies with $A>0.35$ and $A>S$ visually confirmed to be mergers as a function of stellar mass.  The red dashed line represents a constant fit and the green lines represent the $1 \sigma$ confidence intervals.}\label{Vcheck}
\end{figure}

When visually examining the galaxies, a system was considered to be a merger if there were clear signs of tidal debris, tidal tails or bridges, a clearly offset bulge, double nuclei, or generally disturbed and clearly highly asymmetrical morphology.  Galaxies which clearly met these criteria were classified as \emph{Mergers}, while galaxies which were clearly symmetrical, and apparently undisturbed, were classified as \emph{Non-Mergers} (see Figure \ref{maybemerger}).  Galaxies which appeared to be possible mergers but for which a definitive classification could not be made (often due to image noise and low surface brightness) were classified as \emph{Maybe}.  The fraction of true mergers was then found as follows,

\begin{equation}\label{vunseq}
	f_{tm} = \frac{N_{Merger}+N_{Maybe}/2}{N_{Sample}} \pm \sqrt{\frac{f_{tm}\times(1-f_{tm})}{N_{Sample}}},
\end{equation}
where $N_{Sample}$ is the sample size, $N_{\emph{Merger}}$ is the number of galaxies classified as a clear merger, and $N_{\emph{Maybe}}$ is the number of galaxies classified as a possible merger.  
Here we assume half of the galaxies classified as $\emph{Maybe}$ are mergers, although it is likely that most of these galaxies are in fact mergers, as an attempt was made to be conservative in classifying galaxies as a $\emph{Merger}$.  Errors are calculated using the normal approximation of the binomial confidence interval. The fraction of true mergers, $f_{tm}$ is shown as a function of mass in Figure \ref{Vcheck} and the merger classification results are given in Table \ref{vcheckedtable}.  
We find that $f_{tm}$ depends on mass and decreases towards lower masses.
The data in Figure \ref{Vcheck} is fit using the least-squares Marquardt-Levenberg algorithm with a linear fit of the form

\begin{equation}\label{vcheckfiteq}
f_{tm} = m \times \rm{log}_{10}(M_{*}/M_{\odot}) + b, 
\end{equation}
where $m=0.077\pm0.010$ and $b=0.738\pm0.009$.

The average level of contamination between $10^{8.0}<M_{*} < 10^{11.5} M_{\odot}$ is $\sim$70\%, which is similar to the findings of \citet{depr2007} who found that $\sim$20\% of the galaxies they selected as being highly asymmetric were false identifications, although their sample consists of more massive galaxies, which explains their lower contamination level.

\subsection{Separating Major and Minor Mergers}\label{majorminorsep}

In this section we investigate the nature of highly asymmetric galaxies in the nearby universe and determine what fraction of highly asymmetric galaxies are produced through major versus minor mergers.  This is an outstanding question, as we still do not know for certain the mass range which produces highly asymmetric galaxies, which is vital to understand if we are to apply this technique in other samples.

The CAS method is known to primarily identify major mergers, but can also identify minor mergers if the progenitor galaxies' gas fractions are high enough \citep{lotz2010b}. In the later stages of a merger, when the merging galaxies are completely fused, it is difficult to determine the masses of the progenitors. On the other hand, it is possible to know their masses while they are still separated in close pairs.  We will now take a detailed look at the effect that the masses and mass ratios of galaxy pairs have on their asymmetry, as well as determine the separation where paired galaxies become highly asymmetric.  We then use this information to calculate the fraction of highly asymmetric galaxies that are due to major mergers.

\subsubsection{The Dependence of Asymmetry on Pair Separation, Mass, and Mass Ratio}\label{asymmrpmass}

\begin{figure}
\begin{center}
  \begin{overpic}[width=2.2in,angle=270]{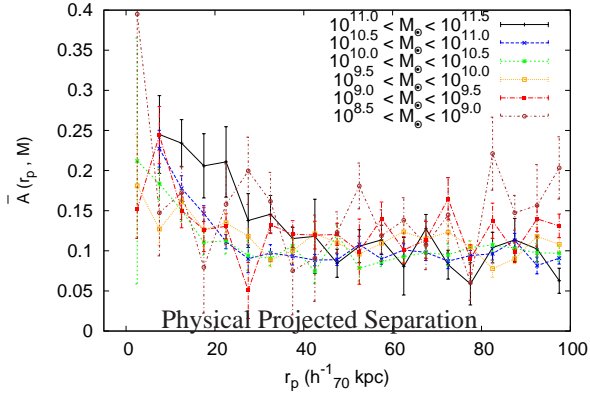}
	\put(60,30){\large Physical Projected Separation} 
  \end{overpic}
\end{center}
\caption{Mean $A$ of major merger pairs versus physical projected separation for a range of masses.  These masses correspond to the masses of the individual pair members, and can be either the heavy or light member.  Higher mass pairs are becoming highly asymmetric at larger physical separations than lower mass pairs.}\label{Aphysical}
\end{figure}

\begin{figure}
\begin{center}
  \begin{overpic}[width=2.2in,angle=270]{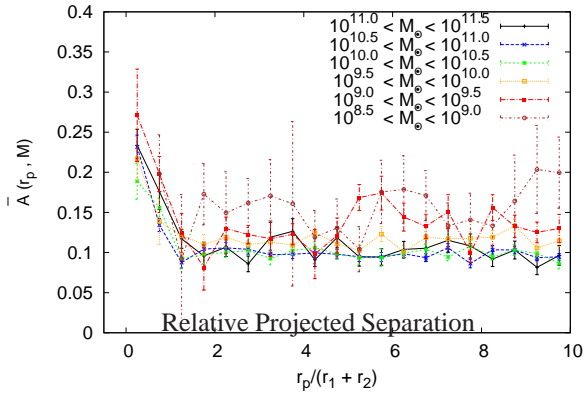}
	\put(60,30){\large Relative Projected Separation} 
  \end{overpic}
\end{center}
  \caption{The mean asymmetry of major merger pairs is plotted versus relative projected separation, where the physical separation is divided by the sum of the Petrosian 90 galaxy radii ($r_p/(r_1+r_2)$).  These masses correspond to the masses of the individual pair members, and can be either the heavy or light member.  All pairs are becoming highly asymmetric as the same relative separation of $r_p<(r_1 + r_2)$.}\label{Arelative}
\end{figure}

\begin{figure}
\begin{center}
  \begin{overpic}[width=2.2in,angle=270]{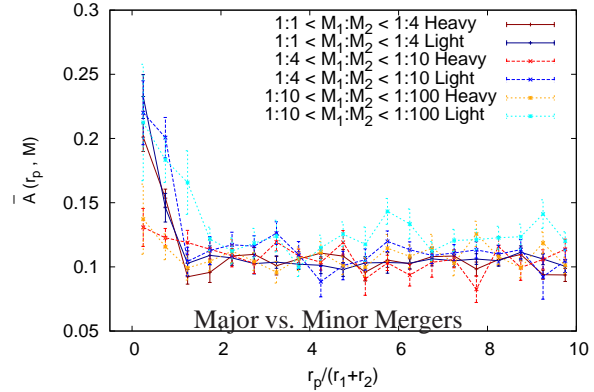}
	\put(70,30){\large Major vs. Minor Mergers} 
  \end{overpic}
\end{center}
  \caption{The mean asymmetry for galaxy pairs of different mass ratios are plotted versus relative projected separation ($r_p/(r_1+r_2)$).  Galaxy pairs are selected from the range $10^{8.0}<M_{*}<10^{11.5} M_{\odot}$, with the red and orange colours representing the heavy member of a pair, and the blue colours representing the lighter member.  In major merger pairs, both members become highly asymmetric for $r_p<(r_1 + r_2)$.  In minor merger pairs the light companion becomes highly asymmetric while the heavy companion does not. Low mass members of minor merger pairs show an increase in mean asymmetry at larger separations, up to $r_p<1.5 \times(r_1 + r_2)$.}\label{HLrelative}
\end{figure}

In order to determine the projected separation where companions produce highly asymmetric galaxies, we look at how the mean asymmetry of dynamically close paired galaxies changes as a function of projected physical and relative separation, mass, and mass ratio.
In \citet{cast2012} it was found that paired galaxies show little signs of interaction for $r_{max} > 120$ $h^{-1}_{70}$ kpc and $\Delta V>500$ km s$^{-1}$.
In order to select pairs which are not interacting, as well as find at what projected separation galaxy pairs begin to become highly asymmetric, pairs are selected up to a projected separation of $r_{max} = 300$ $h^{-1}_{70}$ kpc. A maximum line of sight velocity difference of $\Delta V<500$ km s$^{-1}$ is used to clearly see the transition between strongly interacting and non-interacting galaxies with decreasing separation.

The mean value of $A$ is found for each mass and $r_p$ bin as follows, 

\begin{equation}\label{gamarpsum}
  \bar{A}(r_p,M) = \frac{1}{N} \sum_{i=1}^N{A_i}\;,
\end{equation}
where $N$ is the number of galaxies in a given $r_p$ and mass bin.

In Figure \ref{Aphysical} $\bar{A}$ is shown against physical projected separation for paired galaxies with mass ratios of $<$4:1 for a given mass bin.  Both members of the pair are included in the calculation of $\bar{A}$ and each mass bin includes a mix of heavy and light members from different pairs.  Galaxies in the $10^{11.0}<M_{*}<10^{11.5} M_{\odot}$ bin show an increase in mean asymmetry for $r_{p} < 35$ $h^{-1}_{70}$ kpc, while the $10^{10.5}<M_{*}<10^{11.0} M_{\odot}$ and $10^{10.0}<M_{*}<10^{10.5} M_{\odot}$ bins show increases for $r_{p} < 25$ $h^{-1}_{70}$ kpc and $r_{p} < 20$ $h^{-1}_{70}$ kpc respectively.  This indicates that the high mass major merger pairs are highly asymmetric out to larger physical separations than lower mass pairs. The maximum projected separation identified here agrees with previous work by \citet{depr2007} and \citet{elli2010} who find that the fraction of dynamically close highly asymmetric pairs increases significantly for $r_p < 40$ $h^{-1}_{70}$ kpc.
Although pairs are only plotted for $r_{p} < 100$ $h^{-1}_{70}$ kpc in Figure \ref{Aphysical}, the mean asymmetries of galaxies at all masses remain flat out to 300 $h^{-1}_{70}$ kpc, at which separation the galaxies are essentially isolated.

In Figure \ref{Arelative} the pairs are instead binned by their relative separation (pair separation divided by the sum of the galaxy's Petrosian 90 radii) and we see that the increase in asymmetry is actually occurring at the same relative separation of $r_p < (r_1 + r_2)$ for all galaxy masses. This agrees with the finding by \citet{hern2005} that asymmetry increases (relative to isolated galaxies) for pairs with separations less than $D_{25}$, the photometric diameter of the primary.

In Figure \ref{HLrelative} the mean asymmetry of the heavy and light pair members are shown against relative separation ($r_p/(r_1 + r_2)$) for mass ratio ranges of  $<$4:1, 4:1$-$10:1 and 10:1$-$100:1.  Here all galaxies have been included from the range $10^{8.0}<M_{*}<10^{11.5} M_{\odot}$.
For the major merger pairs (mass ratio of $<$4:1) both the heavy and light companion have an increased mean asymmetry for  $r_p < (r_1 + r_2)$.  As mass ratio increases, light pair members continue to have a high mean asymmetry at small relative separations: $r_p < (r_1 + r_2)$ for the 4:1$-$10:1, and up to $r_p < 1.5 \times(r_1 + r_2)$ for the 10:1$-$100:1 mass ratio pairs.  The heavy members of minor merger pairs also show an increase in mean asymmetry for $r_p < (r_1 + r_2)$, but it is significantly smaller than for major merger pairs.  This result is important in that it shows that the increase in asymmetry for close galaxy pairs is occurring at approximately the same relative separation regardless of mass ratio.

\subsubsection{The Asymmetries of Non-Interacting Projected Pairs}\label{asymcontrol}

\begin{figure}
\begin{center}
\includegraphics[width=2.2in,angle=270]{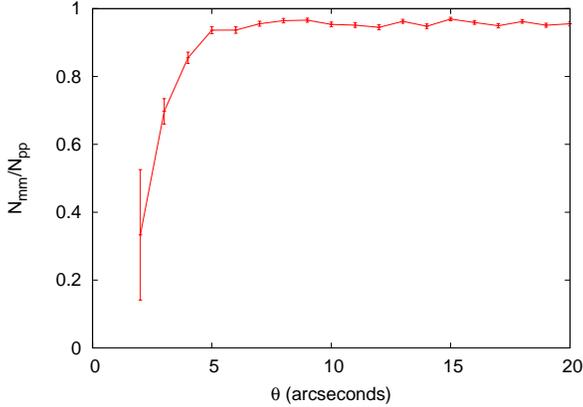}
\end{center}
\caption{The ratio between total number of paired galaxies with mass measurements, $N_{mm}$, and the total number of paired galaxies in the photometric source catalogue, $N_{pp}$, versus angular separation $\theta$.  The errors shown are binomial confidence intervals.}\label{Tcorr}
\end{figure}

The fact that the mean asymmetry of galaxy pairs increases at the same relative separation regardless of mass and mass ratio raises the question of whether this increased asymmetry is simply due to contaminating light from the nearby companion.  To test this we compare the fraction of dynamically close pairs which are highly asymmetric ($\Delta V<500$ km s$^{-1}$) with non-interacting projected pairs  which are highly asymmetric ($1000<\Delta V<100000$ km s$^{-1}$).  For a given range in $\Delta V$ the fraction of pairs which are highly asymmetric is defined as

\begin{equation}\label{pairfraceq}
f_{P_{asym}} = \frac{N_{P_{asym}}}{N_{P_{total}}} = \frac{\sum_{i=1}^{N}{w_{\theta_{asym}}}}{\sum_{i=1}^{N}{w_{\theta_{total}}}} ,
\end{equation}

\noindent where $N_{P_{asym}}$ is the number of paired galaxies in a projected separation bin which satisfy $A>0.35$ and $A>S$ for a given mass ratio range and $N_{P_{total}}$ is the total number of pairs for the same separation bin and mass ratio range.  $w_{\theta}$ is an angular incompleteness weight (see \citealt{patt2000,patt2002}) which we define as

\begin{equation}\label{wtcorr}
w_{\theta} = N_{pp}/N_{mm} ,
\end{equation}

\noindent where $N_{pp}$ is the number of photometric pairs in a given angular separation bin, and $N_{mm}$ is the number of pairs where both galaxies have mass measurements.  Angular separation bins of 1 arcsecond are used and the actual values of $w_{\theta}$ are applied for each separation bin, as opposed to a fit to the incompleteness function.  In Figure \ref{Tcorr} the fraction of photometric pairs where both galaxies have mass measurements ($N_{mm}$/$N_{pp}$) is plotted against angular separation.

In the GAMA survey there is very little dependence of spectroscopic and mass measurement incompleteness on angular separation. Only for angular separations of less than 5 arcseconds does mass incompleteness increase.  Keep in mind that the angular sizes of most galaxies are at least several arcseconds, so this incompleteness only affects galaxies which are essentially overlapping.  Using the raw galaxy counts to calculate $f_{P_{asym}}$ would give a number very close to that given here. The $w_{\theta}$ weight is applied simply to ensure the robustness of the measurement.

\begin{figure}
\begin{center}
  \begin{overpic}[width=2.2in,angle=270]{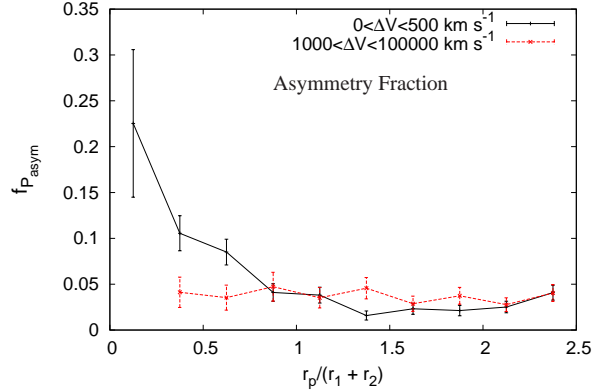}
	\put(100,120){Asymmetry Fraction}
  \end{overpic}
\end{center}
\caption{Fraction of paired galaxies which are highly asymmetric as a function of relative separation for dynamically close pairs ($\Delta V<500$ km s$^{-1}$) and projected pairs ($1000<\Delta V<100000$ km s$^{-1}$).  There are no projected pairs in our sample with $r_p < 0.25 \times (r_1 + r_2)$.}\label{A035control}
\end{figure}

In Figure \ref{A035control} the fraction of paired galaxies which are highly asymmetric is plotted as a function of relative separation for both dynamically close and projected pairs.  Pairs are selected with mass ratios of $<$100:1 and  $10^{8.0}<M_{*}<10^{11.5} M_{\odot}$.  The fraction of highly asymmetric dynamically close pairs increases significantly for $r_p < (r_1 + r_2)$ while the non-interacting projected pairs show no increase with decreasing separation.  This result shows that the increased asymmetry in dynamically close pairs is due to tidally induced morphological disturbances, and not contaminating light from the close companion.

\subsubsection{Fraction of Highly Asymmetric Galaxies in Pairs}

We now look at the fraction of galaxies in close pairs that are highly asymmetric.  
Pairs are selected from the range $10^{6.5}<M_{*}<10^{12.5} M_{\odot}$ to ensure that galaxies in the $10^{8.5}<M_{*}<10^{10.5} M_{\odot}$ range can have companions with mass ratios up to 100:1. There are no galaxies with $M_{*}>10^{12.5} M_{\odot}$ in the GAMA parent sample, meaning no potential $<$100:1 mass ratio pairs are missed for $10^{10.5}<M_{*}<10^{11.5} M_{\odot}$ galaxies.  Of course, due to the increasing mass incompleteness of the sample with increasing redshift, potential low mass companions will be missed.  This incompleteness will affect high mass ratio, minor merger pairs more than major merger pairs.  This effect is probably counteracted by mass dependent clustering, where high mass galaxies are much more likely to have lower mass companions, compared to low mass galaxies. We do not correct for this incompleteness, and as a consequence the fraction of highly asymmetric galaxies which have minor merger companions in the lowest mass bins should be treated as a lower limit.

\begin{figure}
\begin{center}
  \begin{overpic}[width=2.2in,angle=270]{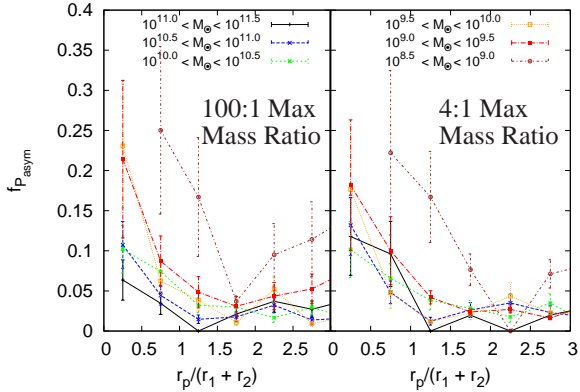}
	\put(75,110){\large 100:1 Max}
	\put(75,100){\large Mass Ratio}
	\put(165,110){\large 4:1 Max}
	\put(165,100){\large Mass Ratio}
  \end{overpic}
\end{center}
  \caption{The fraction of pair members which are highly asymmetric for different relative separation bins and $\Delta V<500$ km s$^{-1}$.  The left panel shows pairs with $<$100:1 mass ratios and the left panel shows pairs with $<$4:1 mass ratios.  The fraction of major merger pairs which are highly asymmetric is fairly constant with mass, while for minor merger pairs there is a strong mass dependence, with a higher fraction of lower mass pair members being highly asymmetric.}\label{Afracall}
\end{figure}

In Figure \ref{Afracall} the fraction of paired galaxies which are highly asymmetric is shown as a function of relative projected separation for different mass bins. 
In the left panel pairs are selected with mass ratios of $<$100:1 and in the right panel with mass ratios of $<$4:1.  A given mass bin can contain either the heavy or light member of different pairs.  As expected, the fraction of paired galaxies that are highly asymmetric increases significantly for $r_p < 1.5 \times(r_1 + r_2)$.  The fraction of highly asymmetric close pairs in the $<$100:1 range also depends on mass, with less massive galaxies having a greater percentage of highly asymmetric close pairs.  Meanwhile the $<$4:1 mass ratio pairs show no apparent trend with mass and have a similar fraction of highly asymmetric galaxies at all masses.  

The highly asymmetric, low mass galaxies in the $<$100:1 range are generally the lower mass members of minor mergers, while the higher mass galaxies are the heavy members that are not experiencing sufficiently strong tidal forces to cause high asymmetry.
These results indicate that high mass galaxies have a greater fraction of minor merger companions than lower mass galaxies.
 This is expected in the hierarchical model of galaxy formation in the $\Lambda CDM$ cosmology (e.g. \citealt{whit1978}), where massive galaxies are built up through mergers and large galaxies are predicted to have numerous smaller companions, which they will eventually merge with due to a loss of orbital momentum through dynamical friction.

\begin{figure}
\begin{center}
  \begin{overpic}[width=2.2in,angle=270]{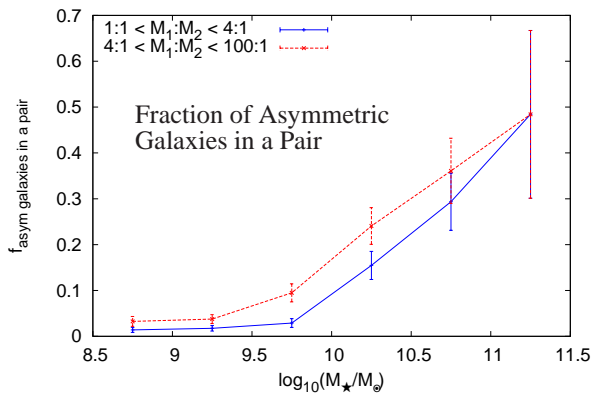}
	\put(50,110){\large Fraction of Asymmetric}
	\put(50,100){\large Galaxies in a Pair}
  \end{overpic}
\end{center}
\caption{The fraction of highly asymmetric galaxies with a close companion ($r_p < (r_1 + r_2)$) as a function of mass.  Higher mass asymmetric galaxies are much more likely to have a close companion than lower mass asymmetric galaxies.}\label{Afracvsmass}
\end{figure}

We now look at what fraction of highly asymmetric galaxies have a close companion with $r_p < 1.5 \times(r_1 + r_2)$.
In Figure \ref{Afracvsmass} the fraction of highly asymmetric galaxies which have a companion is shown as a function of mass.  
Around half of the highest mass asymmetric galaxies have a companion, indicating that about half of the observability time scale of high asymmetry ($T_{merge,A}$) is spent during close passes.  For the lowest mass highly asymmetric galaxies, only about $\sim3\%$ have a companion, indicating that these galaxies spend a greater fraction of $T_{merge,A}$ in merger proper and post merger than higher mass galaxies.  Longer observability time scales are expected for low mass galaxies in the final stages of the merger process due to their much higher gas fractions, as discussed in detail in Section \ref{castimescales}. Additionally, in Section \ref{Vchecksection} the contamination of highly asymmetric galaxies with non-merging systems was found to increase with decreasing mass over the range $10^{8.0}<M_{*}<10^{11.5} M_{\odot}$ from approximately $\sim15\%$ to $\sim40\%$.  These two effects together account for the increase in highly asymmetric paired galaxies with mass.  Note that for most of these pairs, only one of the galaxies is actually highly asymmetric.

 We now determine the number of $<$4:1 mass ratio paired galaxies with $r_p < 1.5 \times(r_1 + r_2)$ where both members are highly asymmetric, $N_{P_{AA}}$.  Using Equation \ref{kappaeq} we calculate the value of $\kappa$, which tells us the average number progenitor galaxies which constitute a typical merger, as detected using the CAS method.  As can be seen from the values in Table \ref{kappatable}, higher mass galaxy pairs are more likely to have both members being highly asymmetric.

\begin{table}
\caption{Values of $\kappa$ as a function of mass, using Equation \ref{kappaeq}.}\label{kappatable}
\centering
\begin{tabular}{c c c c}
\hline\hline
$log(M_{*}/M_{\odot})$ & $N_{A}$ & $N_{P_{AA}}$ & $\kappa$ \\
\hline
8.25 & 72 & 0 & 2.000 \\
8.75 & 248 & 2 & 1.992 \\
9.25 & 399 & 2 & 1.995 \\
9.75 & 279 & 0 & 2.000 \\
10.25 & 201	& 1 & 1.995 \\
10.75 & 102 & 1	& 1.990 \\
11.25 & 19 & 2 & 1.895 \\
\hline\hline
\end{tabular}
\end{table}

\subsubsection{Fraction of Highly Asymmetric Galaxies which are Major Mergers}\label{majormergerfrac}

\begin{figure}
\begin{center}
  \begin{overpic}[width=2.2in,angle=270]{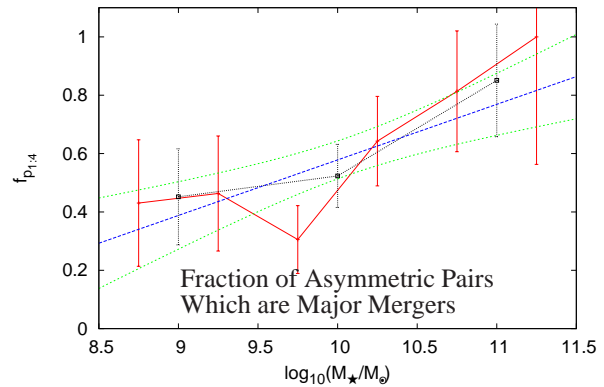}
	\put(65,45){\large Fraction of Asymmetric Pairs}
	\put(65,35){\large Which are Major Mergers}
  \end{overpic}
\end{center}
\caption{The fraction of highly asymmetric $r_p < 1.5 \times(r_1 + r_2)$ pairs which have mass ratios of $<$4:1 is plotted as a function of mass.  The red dashed line represents a linear regression fit to the data and the green dashed lines represent the $1 \sigma$ confidence intervals determined through a Monte Carlo iterative method.}\label{majorfrac}
\end{figure}

\begin{figure}
\begin{center}
  \begin{overpic}[width=2.2in,angle=270]{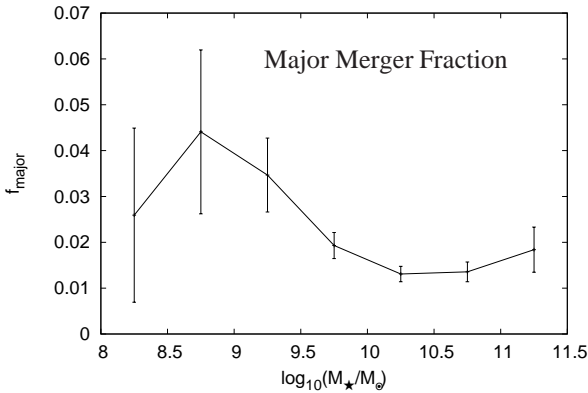}
	\put(100,130){\large Major Merger Fraction}
  \end{overpic}
\end{center}
\caption{The fraction of highly asymmetric galaxies which are on-going major mergers as a function of mass.  For each mass bin the fraction of all highly asymmetric galaxies is weighted to account for only mergers with mass ratios of $<$4:1.}\label{Amajor}
\end{figure}

\begin{table*}
\caption{Values are given here as a function of mass for the total asymmetry fraction, $f_{asym}$, the major merger fraction, $f_{major}$, the major merger rate per galaxy, $R_{major}$, and the co-moving volume major merger rate, $\Gamma_{major}$.}\label{majormergerfracs}
\centering
\begin{tabular}{c c c c c c}
\hline\hline
$log(M_{*}/M_{\odot})$ & $f_{asym}$ & $f_{major}$ & $R_{major}$ & $\Gamma_{major}$ \\
& & & gal$^{-1}$Gyr$^{-1}$ & $h^{3}_{70}$Mpc$^{-3}$Gyr$^{-1}$ \\
\hline
8.25 &	0.0888$\pm$0.0108 &	0.0259$\pm$0.0190 &	0.0125$\pm$0.0092 &	$(3.1\pm2.3)\times10^{-4}$ \\
8.75 &	0.0910$\pm$0.0060 &	0.0441$\pm$0.0179 &	0.0252$\pm$0.0102 &	$(3.7\pm1.5)\times10^{-4}$ \\
9.25 &	0.0524$\pm$0.0027 &	0.0347$\pm$0.0081 &	0.0243$\pm$0.0056 &	$(2.2\pm0.5)\times10^{-4}$ \\
9.75 &	0.0224$\pm$0.0014 &	0.0193$\pm$0.0028 &	0.0173$\pm$0.0025 &	$(1.1\pm0.2)\times10^{-4}$ \\
10.25 &	0.0122$\pm$0.0009 &	0.0131$\pm$0.0017 &	0.0161$\pm$0.0021 &	$(8.5\pm1.1)\times10^{-5}$ \\
10.75 &	0.0105$\pm$0.0010 &	0.0136$\pm$0.0022 &	0.0265$\pm$0.0042 &	$(9.3\pm1.5)\times10^{-5}$ \\
11.25 &	0.0126$\pm$0.0028 &	0.0184$\pm$0.0049 &	0.0456$\pm$0.0122 &	$(2.1\pm5.7)\times10^{-5}$ \\
\hline\hline
\end{tabular}
\end{table*}

The relative number of highly asymmetric galaxies that have a companion within an appropriate projected separation gives us information about the contribution of ongoing major and minor mergers to the highly asymmetric galaxy population.
If we assume that the fraction of highly asymmetric pairs that are major mergers, $f_{P_{4:1}}$, is representative of the fraction of all highly asymmetric galaxies, $f_{A_{4:1}}$ (pairs and isolated) that are major mergers, then we can define

\begin{equation}\label{isopairseq}
f_{A_{4:1}} = f_{P_{4:1}} .
\end{equation}

\noindent Note that in this section we only examine those highly asymmetric galaxies that are in pairs.  While a significant fraction of the highly asymmetric galaxies are not in pairs, the idea here is that the same fraction of minor versus major mergers that produces an asymmetry (as seen in a pair) is similar to the fraction of systems that have already merged.
The fraction of highly asymmetric galaxies in close pairs will
contain two populations: (i) those for which the pair is interacting
and causing the high asymmetry, and (ii) those for which there is an
ongoing merger in addition to the likely future merging of the pair.
Due to the relatively low occurrence of multi-mergers \citep{darg2011}, for small separations most of the galaxies will be from the first population, and $f_{P_{4:1}}$ should provide a good approximation of the fraction of all highly asymmetrical galaxies that are major mergers, $f_{A_{4:1}}$.

For each galaxy that satisfies $A>0.35$ and $A>S$, close companions are identified where $r_p < 1.5 \times(r_1 + r_2)$ and $\Delta V<500$ km s$^{-1}$, with maximum mass ratios of $<$4:1 (major mergers) and $<$100:1 (major and minor mergers). These galaxies are assumed to give the fraction of highly asymmetric paired galaxies that are major mergers and is defined as

\begin{equation}\label{majorfraceq}
f_{P_{4:1}} = \frac{N_{P_{4:1}}}{N_{P_{100:1}}} = \frac{\sum_{i=1}^{N}{w_{\theta}(\text{$<$4:1})}}{\sum_{i=1}^{N}{w_{\theta}(\text{$<$100:1})}} ,
\end{equation}
where $N_{P_{4:1}}$ and $N_{P_{100:1}}$ are the number of paired galaxies in each mass bin that satisfy $A>0.35$ and $A>S$ for the given mass ratio range and $w_{\theta}$ is an angular incompleteness weight defined by Equation \ref{wtcorr}.

The fraction of these pairs that are major mergers is shown in Figure \ref{majorfrac}, where the data is fit with the least-squares Marquardt-Levenberg algorithm with a linear regression of the form

\begin{equation}\label{vfiteq}
f_{P_{4:1}} = m \times \rm{log}_{10}(M_{*}/M_{\odot}) + b, 
\end{equation}
where $m=0.238\pm0.059$ and $b=0.638\pm0.043$.

The major merger companion fraction ($f_{P_{4:1}}$) is highest for more massive galaxies ($\sim90\%$ for the $10^{11.25} M_{\odot}$ bin) and drops progressively for lower mass galaxies ($\sim30\%$ for the $10^{8.75} M_{\odot}$ bin).
Note that almost all of the minor merger galaxies with mass ratios of $>$4:1 identified as being highly asymmetric are the least massive companion.  In Figure \ref{Amajor} the major merger fraction, $f_{major}$, is shown for $10^{8.0}<M_{*}<10^{11.5} M_{\odot}$, with the values given in Table \ref{majormergerfracs}.

\section{Discussion}\label{gamadiscussion}

\subsection{Major Merger Time scales}\label{castimescales}

Galaxy merger time scales are often estimated using dynamical friction (e.g. \citealt{chan1943,whit1976,kitz2008}), with these calculations being valid for small satellite galaxies at relatively large separations.
In Section \ref{asymmrpmass} we found that major merger galaxy pairs begin to become highly asymmetric for $r_p < 1.5 \times(r_1 + r_2)$ over the entire mass range probed.  Since we are dealing with major merger systems that have small relative separations, these calculations are not appropriate.  
As discussed by \citet{hopk2010a}, at small separations orbital energy is lost more through strong resonances between the baryonic components than through dynamical friction.  Also, circular orbits tend to become highly radial as the interaction progresses, leading to shorter merger times.  For these reasons it is best to use N-body simulations to accurately determine the merger time scales of galaxies with very small separations.

Studies using N-body simulations \citep{cons2006,lotz2008b,lotz2010a,lotz2010b} have found that interacting galaxies of similar mass are most asymmetric when they are undergoing a close pass and in the later stages of a merger.  In these simulations it was found that the first peak in asymmetry occurs during the first close pass of an interacting galaxy pair, and then peaks again during the second pass and throughout a significant portion of the post merger.

Simulations by \citet{cons2006} considered galaxies with mass ratios of $<$1:3 and found that the peak in asymmetry during first pass generally lasts around $\sim0.2$ Gyrs.  Note that these simulations do not model the effects of gas.  Depending on the orbital parameters of the interaction, the second close pass occurs 0.6 to 0.8 Gyrs after the first, with the second peak in asymmetry lasting an additional $\sim0.2$ Gyrs. Asymmetry was then found to peak again, or remain high for an additional $\sim0.2$ Gyrs during merger proper.  From these simulations \citet{cons2006} derived the following relation for the average amount of time a galaxy remains highly asymmetric, $T_{merger,A}$, during a merger event,

\begin{equation}\label{conseq}
T_{merger,A} = (0.23\pm0.05)N_{fly} + (0.15\pm0.05)\left(  M_{tot} \over 10^{11} M_{\odot} \right)^{0.25} ,
\end{equation}
where $N_{fly}$ is the number of close passes a galaxy experiences before final merger, and $M_{tot}$ is the total mass of the galaxy.

From the above equation we see that more massive galaxies are highly asymmetric for longer periods during post merger than lower mass galaxies.  The primary galaxies in these merger simulations have total masses of $3.2 \times 10^{11} M_{\odot}$ and stellar masses of $5.98 \times 10^{10} M_{\odot}$.  If we assume that the galaxies in our sample have the same stellar mass to total mass ratio, and that each galaxy undergoes one close pass before final merger, then $T_{merger,A}\sim0.42$ Gyrs for $M_{*} = 10^{11.25} M_{\odot}$ and $T_{merger,A}\sim0.34$ Gyrs for $M_{*} = 10^{10.25} M_{\odot}$.

Meanwhile, the relative amount of gas available in galaxies for star formation increases strongly with decreasing stellar mass (e.g. \citealt{cati2010}), and \citet{lotz2010b} found that gas rich disc galaxies can remain highly asymmetric for significantly longer times.  Using simulations they found that the detection time for 1:3 mass ratio mergers with $A>0.35$ can be approximated by

\begin{equation}\label{gasfraceq}
T_{merger,A}= (-0.26\pm0.05) + (2.28\pm0.23) f_{gas} ,
\end{equation}
where $f_{gas}$ is the relative gas fraction.
In these simulations the primary galaxies have total masses of 1.2$\times 10^{12} M_{\odot}$ and baryonic masses of 6.2$\times 10^{10} M_{\odot}$ with a range of stellar to gas mass ratios.

\citet{cati2010} find that the average $[M_{HI}/M_{*}]$ gas fractions increases from $f_{gas}\sim0.025$ for a $M_{*} = 10^{11.25} M_{\odot}$ galaxy to  $f_{gas}\sim0.32$ for a $M_{*} = 10^{10.25} M_{\odot}$ galaxy.
Using these values in Equation \ref{gasfraceq} we find $T_{merger,A}\sim-0.20$ Gyrs for $M_{*} = 10^{11.25} M_{\odot}$, meaning it is not detectable, and increases to $T_{merger,A}\sim0.46$ Gyrs for $M_{*} = 10^{10.25} M_{\odot}$.  Looking at our major merger fraction measurement in Figure \ref{Amajor} it is clear that merging galaxies with $M_{*} \sim 10^{11.25} M_{\odot}$ are indeed detected using the $A>0.35$ and $A>S$ criteria, implying massive galaxies with very low gas fractions are still being detected using this method.  Therefore the detectability time scale as a function of mass is likely a combination of the asymmetry time scale of the stellar component alone (equation \ref{conseq}) combined with the gas component/star formation asymmetry time scale (equation \ref{gasfraceq}). 
If we assume that we can simply sum these two different estimates of $T_{merger,A}$ to obtain the total detection time scale, then $T_{merger,A}\sim0.42$ Gyrs for $M_{*} = 10^{11.25} M_{\odot}$ and $T_{merger,A}\sim0.80$ Gyrs for $M_{*} = 10^{10.25} M_{\odot}$.

\begin{figure}
\begin{center}
\includegraphics[width=2.2in,angle=270]{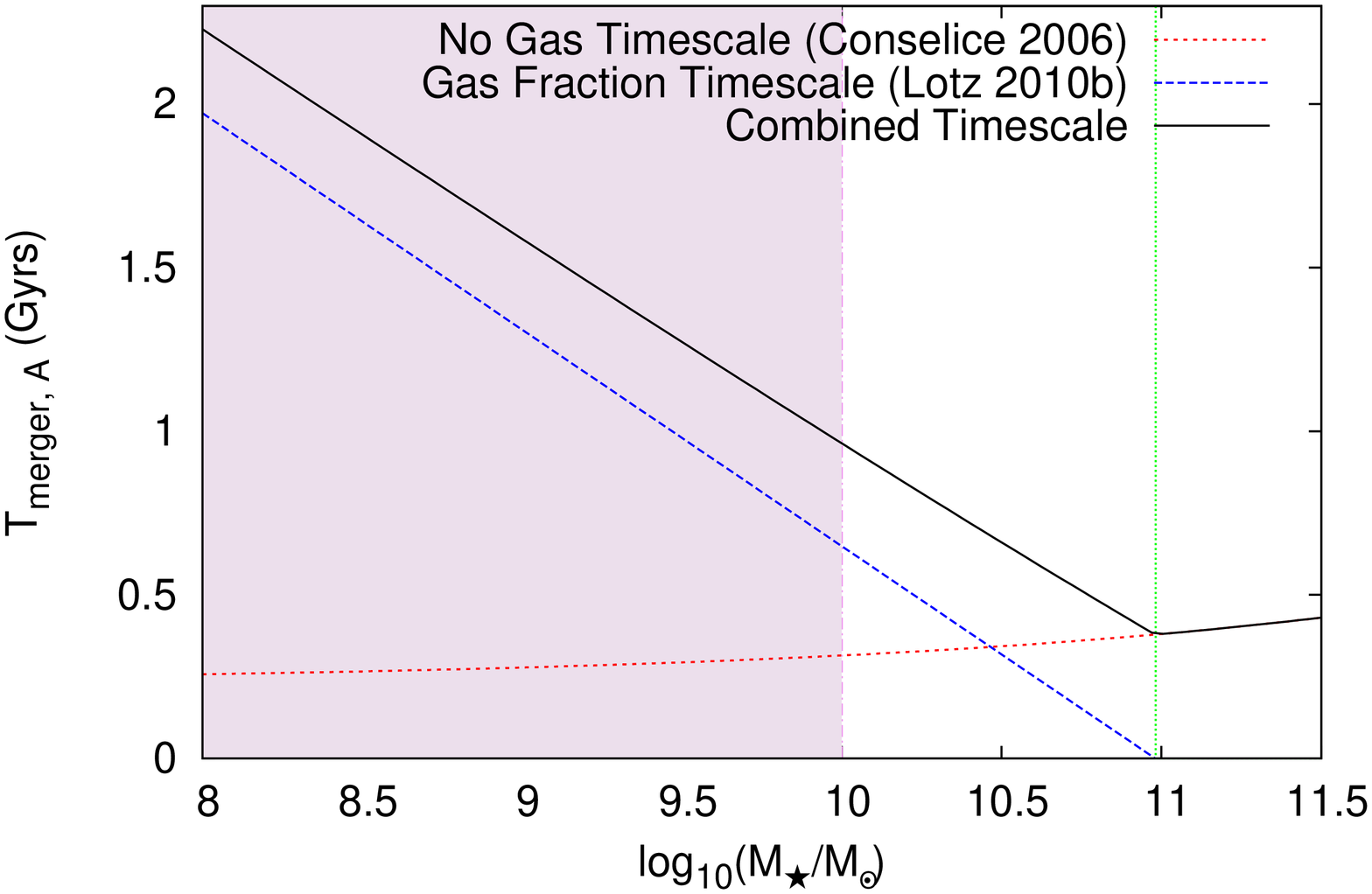}
\end{center}
\caption{The observation time scale models of \citet{cons2006} (red short-dash) and \citet{lotz2010b} (blue long-dash).  The \citet{lotz2010b} relation (Equation \ref{gasfraceq}) is calculated using a linear fit to the \citet{cati2010} mean gas fraction data as a function of mass.  Gas fractions are available for $M_{*} > 10^{10.0} M_{\odot}$, and for lower masses an extrapolation to this fit is used (violet shaded area).  The black solid line represents the combined observability time scale and the vertical green dotted line indicates the mass where the \citet{lotz2010b} time scale goes to zero (as a result of the very low mean gas fraction at this mass).}\label{Tmerger}
\end{figure}

The $T_{merger,A}$ mass dependent relationships of \citet{cons2006} and \citet{lotz2010b} are shown together in Figure \ref{Tmerger}.  The \citet{lotz2010b} relation (equation \ref{gasfraceq}) is calculated using a linear fit to the \citet{cati2010} mean gas fraction data as a function of mass.

These equations are probably over-simplifications of the detection time scales for galaxies of different masses, but they illustrate the fact that the gas fraction of a galaxy has a much greater effect on it being detected with $A>0.35$ than does its mass alone.  Since gas fraction increases strongly with decreasing mass, lower mass galaxies should be detectable in major mergers for significantly longer periods of time than their more massive counterparts.

\begin{figure}
\begin{center}
  \begin{overpic}[width=2.2in,angle=270]{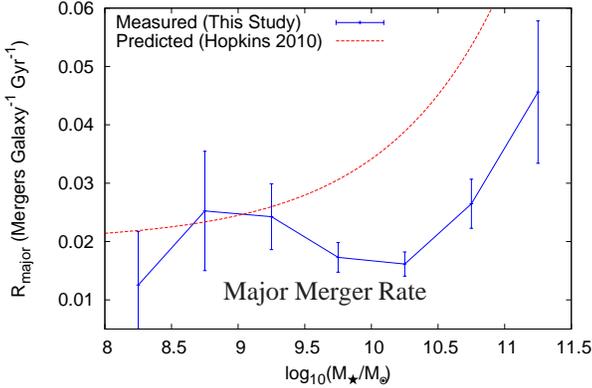}
	\put(83,40){\large Major Merger Rate}
  \end{overpic}
\end{center}
\caption{The galaxy merger rate as a function of stellar mass.  The solid blue line represents the merger rate measured in this work, and the predicted merger rate of \citet{hopk2010b} is represented by the red dashed line.}\label{majormergerrate}
\end{figure}

In Figure \ref{majormergerrate} the galaxy merger rate is shown as a function of mass using using Equation \ref{majormergerrateeq}.  We find that a $M_{*} = 10^{11.25} M_{\odot}$ galaxy experiences $\sim 0.046$ major mergers per Gyr while a $M_{*} = 10^{10.25} M_{\odot}$ galaxy experiences $\sim 0.016$ major mergers per Gyr, implying that the major merger rate approximately triples over this mass range (see Table \ref{majormergerfracs}).  

\begin{figure}
\begin{center}
  \begin{overpic}[width=2.2in,angle=270]{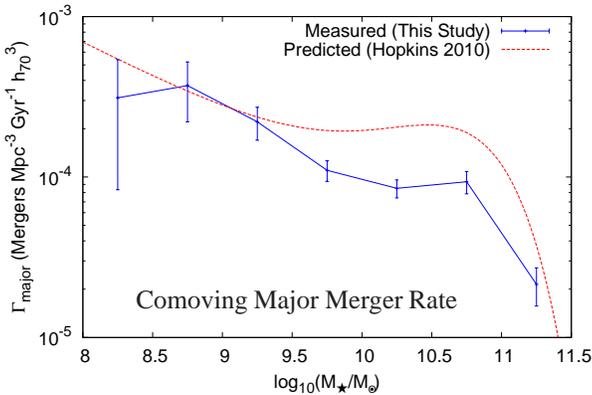}
	\put(50,40){\large Comoving Major Merger Rate}
  \end{overpic}
\end{center}
\caption{The co-moving volume galaxy merger rate as a function of stellar mass.  The solid blue line represents the merger rate measured in this work, and the predicted merger rate of \citet{hopk2010b} is represented by the red dashed line.  The co-moving volume densities used here are from \citet{bald2012}.}\label{comovingrate}
\end{figure}

In Figure \ref{comovingrate} the co-moving volume galaxy merger rate is presented using Equation \ref{comovemergerrateeq}.  The galaxy mass function measurements of \citet{bald2012} are used here, specifically Equation 6 from their paper.
From this we estimate that the major merger rate is $(1.2 \pm 0.5) \times 10^{-3}$ $h^{3}_{70}$ Mpc$^{-3}$ Gyr$^{-1}$ for $10^{8.0}<M_{*}<10^{11.5} M_{\odot}$.

\subsection{Detection of Minor Mergers}

Studies which use N-body simulations and do not consider the effects of changing gas fraction \citep{cons2006,lotz2010a} find that the CAS system is most sensitive to mergers where the galaxies are of similar mass (mass ratios of $<$4:1).  In these simulations minor mergers with greater mass ratios were generally not detected as mergers in CAS, or were for only a very short period of time compared to major mergers.  When the gas fraction of a galaxy is taken into consideration, the picture changes considerably, with \citet{lotz2010b} finding that minor merger events were detectable up to mass ratios of 1:9 or greater.  As discussed in the previous section, galaxy gas fraction increases strongly for lower mass galaxies, and so we would expect minor mergers to be more detectable for lower mass pairs.

In Section \ref{asymmrpmass} we found that with decreasing pair separation the low mass galaxy in a minor merger becomes highly asymmetric while the more massive companion shows little change.  Additionally, the fraction of galaxies identified as being in a major merger with a mass ratios of $<$4:1 decreases progressively towards lower masses.  For $10^{11.0} <M_{*}< 10^{11.5} M_{\odot}$ $\sim85\%$ of the highly asymmetric galaxies with a companion are in a major merger, while for $10^{8.5} <M_{*}< 10^{9.0} M_{\odot}$ this fraction drops to $\sim35\%$.
Of course the smaller companion in a minor merger is usually destroyed by the strong tidal field of the larger galaxy so it is expected that it will become highly asymmetric at some point. The high gas fractions of lower mass galaxies also likely prolongs and enhances the high asymmetry in low mass minor mergers.
A combination of contaminating light from the primary, inherent asymmetry, as well as high gas fractions, are likely combining to produce a high asymmetry measurement in these galaxies.

\subsection{Comparison with Other Studies and Theory}

In Section \ref{majormergerfrac} we found that the major merger fraction is fairly constant at $\sim1.3-2\%$ between $10^{9.5}<M_{*}<10^{11.5} M_{\odot}$, and increases to $\sim4\%$ for lower masses.  This is consistent with the result of \citet{brid2010} who find a decreasing merger fraction with increasing mass using visually selected mergers, although the large errors on our lower mass data make our measurement uncertain for $M_{*}<10^{9.5} M_{\odot}$.
For $10^{9.5}<M_{*}<10^{11.5} M_{\odot}$ our results agrees well with the low redshift pair fractions of \citet{patt2008}, \citet{domi2009} and \citet{xu2012}, who all find a relatively constant major merger fraction as a function of mass (or luminosity).  
In Section \ref{castimescales} we argued that lower mass major mergers should be detectable for significantly longer periods than higher mass major mergers, due to the strong increase in gas fraction with decreasing mass.  Therefore we expect the actual major merger rate at low redshift to increase with mass.  Pair fractions are susceptible to interloper galaxies and \citet{patt2008} used an N-body simulation to determine that this contamination increases strongly with decreasing luminosity.  If the pair fractions mentioned above were corrected for interloper contamination, they would all show a positive increase with mass.
The technique used in this paper should be much less susceptible to interlopers due to the use of morphological information.

The fraction of wet mergers is known to increase with redshift (e.g. \citealt{lin2008}) so galaxies equivalent to the high mass galaxies probed in this study would have been more gas rich in the past.  At higher redshift one would expect that the detectability  time scale of massive galaxies using the CAS method should be longer, resulting in a positively increasing merger fraction with mass, as found by \citet{cons2008}.  The studies of \citet{xu2004} and \citet{bund2009} both find evidence that the pair fraction increases positively with mass, although the errors in these studies are rather large.  These observations support the findings of \citet{hopk2010b} who found using semi-empirical models that the absolute merger rates at a fixed mass ratio increase with galaxy mass.  

In Figures \ref{majormergerrate} and \ref{comovingrate} the merger rate per galaxy and the co-moving volume galaxy merger rates are compared to the model predictions of \citet{hopk2010b} (for $z=0$).  From these figures it is clear that our measurements generally agree with the model predictions, especially for the lower mass galaxies, however, the models predict a slightly steeper increase in the galaxy merger rate with increasing mass compared to our results.
The simulations of \citet{lotz2010b} and \citet{cons2006} used only disk galaxies to calculate the values of $T_{merger,A}$, while in reality, early type, spheroidal systems dominate the populations of higher mass galaxies.  It is possible that $T_{merger,A}$ is significantly shorter for mergers between spheroidal galaxies compared to disk galaxies, resulting in higher merger rates, particularly for higher mass galaxy populations.  For this reason, and because of the difficulty in determining the morphological types of progenitor galaxies in advanced mergers, no effort was made in this study to sub-divide the galaxy merger rates based on morphology.
Future work determining what effect the progenitor galaxy's morphologies have on $T_{merger,A}$ is clearly needed.   In light of this, the merger rates presented here are likely lower limits, especially for higher mass, spheroidally dominated populations.

As mentioned previously, galaxies selected using the CAS method are predominantly undergoing a close pass or in post merger.  Therefore the merger fraction obtained using this method misses most of the galaxies which are between close passes and at large relative separation, although our estimate of the major merger \emph{rate} accounts for these missed galaxies statistically.  In \citet{cast2012} it was found that galaxies with very loose tidal arms can be identified with a companion up to $\sim 120$ $h^{-1}_{70}$ kpc.  Similarly, \citet{patt2013} find evidence of interaction enhanced star formation up to $\sim 150$ $h^{-1}_{70}$ kpc.  Such galaxies are likely near their maximum separation between their first and second close passes.  In order to obtain a definitive measurement of the mass dependent merger rate a combination of different methods will need to be used to identify all merging galaxies, at all stages of the merger process.  Future pair studies will need to accurately account for interloper pairs as a function of mass, while studies which use morphological methods, such as this one, will need to correctly estimate the effect which gas fraction, mass and morphology have on the merger detectability time scales.

\section{Conclusions}\label{gamaconclusions}

We study 1470 highly asymmetric galaxies, 142 of which are in very close pairs, as seen in the combination of GAMA data and SDSS imaging.  We analyse these data together to derive properties of the nearby merging population.  Our major results include:

\begin{enumerate}

\item Galaxy pairs at all masses and mass ratios are found to show a significant increase in asymmetry for $r_p < 1.5\times(r_1 + r_2)$.  In major mergers with mass ratios of $<$4:1, both galaxies in a pair show a strong increase in asymmetry, while for minor mergers with mass ratios greater than 4:1, the lower mass companion becomes highly asymmetric, while the larger galaxy is much less affected.

\item The fraction of highly asymmetric galaxies identified as being in a major merger pair (with mass ratios of $<$4:1) decreases progressively towards lower masses.  For $10^{11.0} <M_{*}< 10^{11.5} M_{\odot}$ $\sim85\%$ of the highly asymmetric galaxies with a companion are in a major merger, while for $10^{8.5} <M_{*}< 10^{9.0} M_{\odot}$ this fraction drops to $\sim35\%$.

\item We find that the major merger fraction is fairly constant at $\sim1.3-2\%$ between $10^{9.5}<M_{*}<10^{11.5} M_{\odot}$, and increases to $\sim4\%$ at lower masses.  We argue that lower mass major mergers should be detectable for significantly longer periods than higher mass major mergers, due to the strong increase in galaxy gas fraction with decreasing mass.  The major merger rate is found to increase with increasing mass, approximately tripling over the mass range probed, rising from 0.013$\pm$0.009 mergers galaxy$^{-1}$ Gyr$^{-1}$ at $M_{*} \sim 10^{8.25} M_{\odot}$ to 0.046$\pm$0.012 mergers galaxy$^{-1}$ Gyr$^{-1}$ at $M_{*} \sim 10^{11.25} M_{\odot}$  The total co-moving volume major merger rate for $10^{8.0}<M_{*}<10^{11.5} M_{\odot}$ is calculated as $(1.2 \pm 0.5) \times 10^{-3}$ $h^{3}_{70}$ Mpc$^{-3}$ Gyr$^{-1}$.

\end{enumerate}

\section*{Acknowledgments}

KRVS would like to thank the Science and Technology Facilities Council (STFC) for providing funding for this project, as well as the Government of Catalonia for a research travel grant (ref. 2010 BE-00268) to begin this project at the University of Nottingham.  PN acknowledges the support of the Royal Society through the award of a University Research Fellowship and the European Research Council, through receipt of a Starting Grant (DEGAS-259586).

\bsp
\label{lastpage}

\begin{thebibliography}{99}




\bibitem[\protect\citeauthoryear{Baldry et al.}{2010}]{bald2010} 
Baldry I.~K., et al., 2010, MNRAS, 404, 86 

\bibitem[\protect\citeauthoryear{Baldry et al.}{2012}]{bald2012} 
Baldry I.~K., et al., 2012, MNRAS, 421, 621 


\bibitem[\protect\citeauthoryear{Bamford et al.}{2009}]{bamf2009} Bamford S.~P., et al., 2009, MNRAS, 393, 
1324 

\bibitem[\protect\citeauthoryear{Barnes 
\& Hernquist}{1991}]{barn1991} Barnes J.~E., Hernquist L.~E., 1991, ApJ, 370, L65 


\bibitem[\protect\citeauthoryear{Barnes 
\& Hernquist}{1996}]{barn1996} Barnes J.~E., Hernquist L., 1996, ApJ, 471, 115 


\bibitem[\protect\citeauthoryear{Benson et al.}{2002}]{bens2002} 
Benson A.~J., Lacey C.~G., Baugh C.~M., Cole S., Frenk C.~S., 2002, MNRAS, 
333, 156 

\bibitem[\protect\citeauthoryear{Bertin 
\& Arnouts}{1996}]{bert1996} Bertin E., Arnouts S., 1996, A\&AS, 117, 393 


\bibitem[\protect\citeauthoryear{Bluck et al.}{2012}]{bluc2012} 
Bluck A.~F.~L., Conselice C.~J., Buitrago F., Gr{\"u}tzbauch R., Hoyos C., 
Mortlock A., Bauer A.~E., 2012, ApJ, 747, 34 

\bibitem[\protect\citeauthoryear{Bridge, Carlberg, 
\& Sullivan}{2010}]{brid2010} Bridge C.~R., Carlberg R.~G., Sullivan M., 2010, ApJ, 709, 1067 



\bibitem[\protect\citeauthoryear{Bundy et al.}{2009}]{bund2009} 
Bundy K., Fukugita M., Ellis R.~S., Targett T.~A., Belli S., Kodama T., 
2009, ApJ, 697, 1369 


\bibitem[\protect\citeauthoryear{Casteels et 
al.}{2013}]{cast2012} Casteels K.~R.~V., et al., 2013, MNRAS, 
429, 1051 

\bibitem[\protect\citeauthoryear{Catinella et 
al.}{2010}]{cati2010} Catinella B., et al., 2010, MNRAS, 403, 
683 

\bibitem[\protect\citeauthoryear{Cole et al.}{2000}]{cole2000} 
Cole S., Lacey C.~G., Baugh C.~M., Frenk C.~S., 2000, MNRAS, 319, 168 


\bibitem[\protect\citeauthoryear{Colless et 
al.}{2001}]{coll2001} Colless M., et al., 2001, MNRAS, 328, 1039 


\bibitem[\protect\citeauthoryear{Conselice, Bershady, 
\& Jangren}{2000}]{cons2000} Conselice C.~J., Bershady M.~A., Jangren A., 2000, ApJ, 529, 886 

\bibitem[\protect\citeauthoryear{Conselice}{2003}]{cons2003b} 
Conselice C.~J., 2003, ApJS, 147, 1 



\bibitem[\protect\citeauthoryear{Conselice}{2006}]{cons2006} 
Conselice C.~J., 2006, ApJ, 638, 686 


\bibitem[\protect\citeauthoryear{Conselice, Rajgor, 
\& Myers}{2008}]{cons2008} Conselice C.~J., Rajgor S., Myers R., 2008, MNRAS, 386, 909 


\bibitem[\protect\citeauthoryear{Cox et al.}{2008}]{cox2008} 
Cox T.~J., Jonsson P., Somerville R.~S., Primack J.~R., Dekel A., 2008, 
MNRAS, 384, 386 

\bibitem[\protect\citeauthoryear{Chandrasekhar}{1943}]{chan1943} 
Chandrasekhar S., 1943, ApJ, 97, 255 


\bibitem[\protect\citeauthoryear{Darg et al.}{2011}]{darg2011} 
Darg D.~W., Kaviraj S., Lintott C.~J., Schawinski K., Silk J., Lynn S., 
Bamford S., Nichol R.~C., 2011, MNRAS, 416, 1745 

\bibitem[\protect\citeauthoryear{De Propris et 
al.}{2007}]{depr2007} De Propris R., Conselice C.~J., Liske J., 
Driver S.~P., Patton D.~R., Graham A.~W., Allen P.~D., 2007, ApJ, 666, 212 

\bibitem[\protect\citeauthoryear{De Propris et 
al.}{2014}]{depr2014} De Propris R., et al., 2014, arXiv, 
arXiv:1407.4996 

\bibitem[\protect\citeauthoryear{Di Matteo, Springel, 
\& Hernquist}{2005}]{dima2005} Di Matteo T., Springel V., Hernquist L., 2005, Natur, 433, 604 

\bibitem[\protect\citeauthoryear{Di Matteo et 
al.}{2012}]{dima2012} Di Matteo T., Khandai N., DeGraf C., Feng 
Y., Croft R.~A.~C., Lopez J., Springel V., 2012, ApJ, 745, L29 

\bibitem[\protect\citeauthoryear{Driver et 
al.}{2009}]{driv2009} Driver S.~P., et al., 2009, A\&G, 50, 050000 


\bibitem[\protect\citeauthoryear{Driver et al.}{2011}]{driv2011} 
Driver S.~P., et al., 2011, MNRAS, 413, 971 


\bibitem[\protect\citeauthoryear{Domingue et 
al.}{2009}]{domi2009} Domingue D.~L., Xu C.~K., Jarrett T.~H., 
Cheng Y., 2009, ApJ, 695, 1559 


\bibitem[\protect\citeauthoryear{Ellison et 
al.}{2010}]{elli2010} Ellison S.~L., Patton D.~R., Simard L., 
McConnachie A.~W., Baldry I.~K., Mendel J.~T., 2010, MNRAS, 407, 1514 


\bibitem[\protect\citeauthoryear{Ellison et 
al.}{2013}]{elli2013} Ellison S.~L., Mendel J.~T., Scudder 
J.~M., Patton D.~R., Palmer M.~J.~D., 2013, MNRAS, 430, 3128 

\bibitem[\protect\citeauthoryear{Fakhouri 
\& Ma}{2008}]{fakh2008} Fakhouri O., Ma C.-P., 2008, MNRAS, 386, 577 

\bibitem[\protect\citeauthoryear{Fakhouri, Ma, 
\& Boylan-Kolchin}{2010}]{fakh2010} Fakhouri O., Ma C.-P., Boylan-Kolchin M., 2010, MNRAS, 406, 2267 

\bibitem[Hern{\'a}ndez-Toledo et al.(2005)]{hern2005} 
Hern{\'a}ndez-Toledo, H.~M., Avila-Reese, V., Conselice, C.~J., 
\& Puerari, I.\ 2005, AJ, 129, 682 

\bibitem[\protect\citeauthoryear{Hill et al.}{2011}]{hill2011} 
Hill D.~T., et al., 2011, MNRAS, 412, 765 

\bibitem[\protect\citeauthoryear{Hopkins et 
al.}{2005a}]{hopk2005a} Hopkins P.~F., Hernquist L., Martini P., 
Cox T.~J., Robertson B., Di Matteo T., Springel V., 2005, ApJ, 625, L71 

\bibitem[\protect\citeauthoryear{Hopkins et 
al.}{2005b}]{hopk2005b} Hopkins P.~F., Hernquist L., Cox T.~J., Di 
Matteo T., Martini P., Robertson B., Springel V., 2005, ApJ, 630, 705 

\bibitem[\protect\citeauthoryear{Hopkins et 
al.}{2010a}]{hopk2010a} Hopkins P.~F., et al., 2010, ApJ, 724, 915 


\bibitem[\protect\citeauthoryear{Hopkins et 
al.}{2010b}]{hopk2010b} Hopkins P.~F., et al., 2010, ApJ, 715, 202 

\bibitem[\protect\citeauthoryear{Hopkins et 
al.}{2013}]{hopk2013} Hopkins A.~M., et al., 2013, MNRAS, 430, 
2047 

\bibitem[\protect\citeauthoryear{Jogee et al.}{2009}]{joge2009} 
Jogee S., et al., 2009, ApJ, 697, 1971 


\bibitem[\protect\citeauthoryear{Kauffmann, White, 
\& Guiderdoni}{1993}]{kauf1993} Kauffmann G., White S.~D.~M., Guiderdoni B., 1993, MNRAS, 264, 201 

\bibitem[\protect\citeauthoryear{Kaviraj}{2014}]{kavi2014} 
Kaviraj S., 2014, MNRAS, 437, L41 

\bibitem[\protect\citeauthoryear{Kitzbichler 
\& White}{2008}]{kitz2008} Kitzbichler M.~G., White S.~D.~M., 2008, MNRAS, 391, 1489 

\bibitem[\protect\citeauthoryear{Le F{\`e}vre et 
al.}{2000}]{lefe2000} Le F{\`e}vre O., et al., 2000, MNRAS, 311, 
565 


\bibitem[\protect\citeauthoryear{Lee et al.}{2012}]{lee2012} 
Lee G.-H., Park C., Lee M.~G., Choi Y.-Y., 2012, ApJ, 745, 125 

\bibitem[\protect\citeauthoryear{Lin et al.}{2004}]{lin2004} 
Lin L., et al., 2004, ApJ, 617, L9 

\bibitem[\protect\citeauthoryear{Lin et al.}{2008}]{lin2008} 
Lin L., et al., 2008, ApJ, 681, 232 

\bibitem[\protect\citeauthoryear{Liske et al.}{2003}]{lisk2003} 
Liske J., Lemon D.~J., Driver S.~P., Cross N.~J.~G., Couch W.~J., 2003, 
MNRAS, 344, 307 


\bibitem[\protect\citeauthoryear{L{\'o}pez-Sanjuan et 
al.}{2011}]{lope2011} L{\'o}pez-Sanjuan C., et al., 2011, A\&A, 530, A20 



\bibitem[\protect\citeauthoryear{Lotz et al.}{2008}]{lotz2008} 
Lotz J.~M., et al., 2008, ApJ, 672, 177 

\bibitem[\protect\citeauthoryear{Lotz et al.}{2008}]{lotz2008b} 
Lotz J.~M., Jonsson P., Cox T.~J., Primack J.~R., 2008, MNRAS, 391, 1137 

\bibitem[\protect\citeauthoryear{Lotz et al.}{2010a}]{lotz2010a} 
Lotz J.~M., Jonsson P., Cox T.~J., Primack J.~R., 2010, MNRAS, 404, 575 

\bibitem[\protect\citeauthoryear{Lotz et al.}{2010b}]{lotz2010b} 
Lotz J.~M., Jonsson P., Cox T.~J., Primack J.~R., 2010, MNRAS, 404, 590 


\bibitem[Lotz et al.(2011)]{lotz2011} Lotz, J.~M., Jonsson, P., 
Cox, T.~J., et al.\ 2011, ApJ, 742, 103 


\bibitem[\protect\citeauthoryear{M{\'e}ndez-Hern{\'a}ndez et 
al.}{2011}]{mend2011} M{\'e}ndez-Hern{\'a}ndez H., Maga{\~n}a 
A.~M., Hern{\'a}ndez-Toledo H.~M., Valenzuela O., 2011, RMxAC, 40, 78 


\bibitem[\protect\citeauthoryear{Mihos 
\& Hernquist}{1994}]{miho1994} Mihos J.~C., Hernquist L., 1994, ApJ, 431, L9 

\bibitem[\protect\citeauthoryear{Mihos 
\& Hernquist}{1996}]{miho1996} Mihos J.~C., Hernquist L., 1996, ApJ, 464, 641 


\bibitem[\protect\citeauthoryear{Patton et al.}{2000}]{patt2000} 
Patton D.~R., Carlberg R.~G., Marzke R.~O., Pritchet C.~J., da Costa L.~N., 
Pellegrini P.~S., 2000, ApJ, 536, 153 

\bibitem[\protect\citeauthoryear{Patton et al.}{2002}]{patt2002} 
Patton D.~R., et al., 2002, ApJ, 565, 208 

\bibitem[\protect\citeauthoryear{Patton et al.}{2005}]{patt2005} 
Patton D.~R., Grant J.~K., Simard L., Pritchet C.~J., Carlberg R.~G., Borne 
K.~D., 2005, AJ, 130, 2043 

\bibitem[\protect\citeauthoryear{Patton 
\& Atfield}{2008}]{patt2008} Patton D.~R., Atfield J.~E., 2008, ApJ, 685, 235 


\bibitem[\protect\citeauthoryear{Patton et al.}{2013}]{patt2013} 
Patton D.~R., Torrey P., Ellison S.~L., Mendel J.~T., Scudder J.~M., 2013, 
arXiv, arXiv:1305.1595 

\bibitem[\protect\citeauthoryear{Petrosian}{1976}]{petr1976} 
Petrosian V., 1976, ApJ, 209, L1 

\bibitem[\protect\citeauthoryear{Robotham et 
al.}{2014}]{robo2014} Robotham A.~S.~G., et al., 2014, arXiv, 
arXiv:1408.1476 

\bibitem[\protect\citeauthoryear{Skibba et al.}{2012}]{skib2011} 
Skibba R.~A., et al., 2012, MNRAS, 423, 1485 

\bibitem[\protect\citeauthoryear{Tasca et 
al.}{2014}]{tasc2014} Tasca L.~A.~M., et al., 2014, A\&A, 565, A10 


\bibitem[Taylor-Mager et al.(2007)]{tayl2007} Taylor-Mager, 
V.~A., Conselice, C.~J., Windhorst, R.~A., 
\& Jansen, R.~A.\ 2007, ApJ, 659, 162 


\bibitem[\protect\citeauthoryear{Taylor et al.}{2011}]{tayl2011} 
Taylor E.~N., et al., 2011, MNRAS, 418, 1587 

\bibitem[\protect\citeauthoryear{Torrey et al.}{2014}]{torr2014} 
Torrey P., Vogelsberger M., Genel S., Sijacki D., Springel V., Hernquist 
L., 2014, MNRAS, 438, 1985 


\bibitem[\protect\citeauthoryear{White}{1976}]{whit1976} White 
S.~D.~M., 1976, MNRAS, 174, 467 

\bibitem[\protect\citeauthoryear{White 
\& Rees}{1978}]{whit1978} White S.~D.~M., Rees M.~J., 1978, MNRAS, 183, 341 


\bibitem[Xu et al.(2004)]{xu2004} Xu, C.~K., Sun, Y.~C., \& 
He, X.~T.\ 2004, ApJl, 603, L73 


\bibitem[\protect\citeauthoryear{Xu et al.}{2012}]{xu2012} Xu 
C.~K., Zhao Y., Scoville N., Capak P., Drory N., Gao Y., 2012, ApJ, 747, 85 


\bibitem[\protect\citeauthoryear{York et al.}{2000}]{york2000} 
York D.~G., et al., 2000, AJ, 120, 1579 








\end{thebibliography}
\end{document}